\documentstyle[12pt,a41,cite,amssymb]{article}

\newcommand{\C}{{\mathbb C}}
\newcommand{\Li}{{\rm Li}}

\newcommand{\bq}{\begin{equation}}
\newcommand{\eq}{\end{equation}}
\newcommand\beq{\begin{equation}}
\newcommand\eeq{\end{equation}}
\newcommand\bea{\begin{eqnarray}}
\newcommand\eea{\end{eqnarray}}

\newcommand\Mvec{\,\mbox{\bf M}}

\include{epsfig}

\begin{document}
\noindent
\sloppy

\thispagestyle{empty}
\begin{flushleft}
DESY 06--040 \hfill
{\tt hep-ph/0604019}\\
SFB/CPP-06-14\\
March 2006
\end{flushleft}
%
%\setcounter{page}{0}
% 1
%\mbox{}
\vspace*{\fill}
\begin{center}
{\boldmath\Large\bf $O(\alpha_s^2)$ Timelike Wilson Coefficients for}

\vspace{2mm}
{\Large\bf Parton--Fragmentation  Functions in Mellin Space}

\vspace{2cm}
\large
Johannes Bl\"umlein$^a$  and
Vajravelu  Ravindran$^{a,b}$
\\
\vspace{2em}
\normalsize
{\it $^a$~Deutsches Elektronen--Synchrotron, DESY,\\
Platanenallee 6, D--15738 Zeuthen, Germany}
\\

\vspace{2mm}
{\it $^b$~Harish--Chandra Research Institute, Chhatnag Road,\\
 Jhunsi, Allahabad, India.}
\\
\vspace{2em}
%%\today
\end{center}
\vspace*{\fill}
%
%%%%%%%%%%%%%%%%%%%%%%%%%%%%%%%%%%%%%%%%%%%%%%%%%%%%%%%%%%%%%%%%%%%%%%%%
\begin{abstract}
\noindent
We calculate the Mellin moments of the $O(\alpha_s^2)$ coefficient functions 
for the unpolarized and polarized fragmentation functions. They can be expressed
in terms of multiple finite harmonic sums of maximal weight {\sf w = 4}. Using 
algebraic and structural relations between harmonic sums one finds that besides 
the single harmonic sums only three basic sums and their derivatives w.r.t. the 
summation index contribute. The Mellin moments are analytically continued to
complex values of the Mellin variable. This representation significantly 
reduces the large complexity being present in $x$--space calculations and allow 
very compact and fast numerical implementations.
\end{abstract}
%%%%%%%%%%%%%%%%%%%%%%%%%%%%%%%%%%%%%%%%%%%%%%%%%%%%%%%%%%%%%%%%%%%%%%%%
\vspace*{\fill}
\newpage
%%%%%%%%%%%%%%%%%%%%%%%%%%%%%%%%%%%%%%%%%%%%%%%%%%%%%%%%%%%%%%%%%%%%%%%%%%%%%
\section{Introduction}
%%%%%%%%%%%%%%%%%%%%%%%%%%%%%%%%%%%%%%%%%%%%%%%%%%%%%%%%%%%%%%%%%%%%%%%%%%%%%

\vspace{1mm}\noindent
The production of hadrons in annihilation processes such as $e^+ e^-$ or 
quark-antiquark and gluon-gluon scattering is described by parton--hadron 
fragmentation in the final state of these processes. Due to the factorization 
theorems perturbative Quantum Chromodynamics describes the hard scattering 
sub-processes at the level of leading twist, while hadron formation requires 
non--perturbative functions, the fragmentation densities $D_p^H(z,M^2)$. At leading 
order in the coupling constant they describe the probability that a final state 
parton $p$ involved in the hard scattering process produces a hadron $H$. Here 
$z$ denotes the momentum fraction of the parton in the hadron and $M^2$ is 
the respective factorization scale. At higher orders the fragmentation process is 
described by fragmentation functions which are formed of Mellin convolutions 
of the fragmentation densities and the corresponding time--like Wilson coefficients.

The process of $e^+e^-$ annihilation into hadrons provides the cleanest environment
to measure the fragmentation functions and to extract the individual fragmentation 
densities~:
%---------------------------------------------------------------------------------
\begin{equation}
e^+ + e^- \rightarrow \overline{H} + X~.
\end{equation}
%---------------------------------------------------------------------------------
By crossing from $s$-- to $t$--channel this process  transforms into deeply 
inelastic scattering of an electron off the hadron $H$~:
%---------------------------------------------------------------------------------
\begin{equation}
e^- + H \rightarrow e^- + X~.
\end{equation}
%---------------------------------------------------------------------------------
Both processes are related by this transformation and the question arises for
similarities or transformation properties of characteristic quantities contributing
to each of the reactions. This question has been raised early \cite{DLY1} concerning
both the non--perturbative and perturbative components. 
Crossing of the non--perturbative fragmentation- and structure functions is not expected to 
hold  \cite{CROS}. However, one may ask the question for the perturbative part
order by order in the strong coupling constant $a_s = \alpha_s/(4\pi)$.  
Here it is not expected that the crossing holds for the Wilson coefficients and splitting 
functions defined in a specific factorization scheme, as e.g. the $\overline{\rm MS}$ 
scheme, individually. However, physical quantities as physical evolution kernels $K_{ij}$ obey 
the crossing relations. These functions describe the scale evolution of observables and
are scheme--independent. To NLO they were derived in 
\cite{Furmanski:1981cw,Grunberg:1982fw,Catani:1996sc,BNR}
for polarized and unpolarized targets (final state hadrons) referring to different observables as 
the structure (fragmentation) functions and their slope and combining the longitudinal and 
transverse projections $F_2$ and $F_L$. Recently the scheme-invariant evolution kernels were 
derived at NNLO in \cite{Blumlein:2004xs} for the space--like  case\footnote{The structure in the 
time--like case is analogous.}. In Ref.~\cite{BNR} the conditions 
were derived up to the 2--loop Wilson coefficients. The crossing of the 2-loop splitting functions
was studied \cite{Stratmann:1996hn} based on Refs.~\cite{CF}.

To be capable to construct physical evolution kernels in the space-- and time--like
cases, the complexity of the respective integrals requires to choose the Mellin space
representation. For the whole set of the time--like Wilson coefficients so far only
the $x$--space representation has been 
available~\cite{Rijken:1996ns,Rijken:1996np,Rijken:1996vr,Rijken:1997rg}\footnote{In the 
present analysis typos contained in the print were accounted for.}. In this paper 
we 
derive the Mellin--representation of the 2--loop Wilson coefficients. As observed
in case of other single--scale quantities at the 2-- and 3--loop level 
before \cite{Blumlein:2004bb,Blumlein:2004bc,Blumlein:2005im,Blumlein:2005jg,JBSMx} 
a considerable structural simplification for these quantities
is obtained w.r.t. the number of contributing functions. In the Mellin-space representation  
the evolution equations can be solved analytically and fast and precise numerical 
implementations are obtained \cite{Gluck:1991ee,Blumlein:1997em}. 
The 2--loop  Wilson coefficients depend only on a few universal basic functions, for which 
representations in the complex Mellin-plane have to be derived.

The paper is organized as follows. In section~2 the differential cross sections for the process
$e^+ e^- \rightarrow H + X$ are summarized and expressed in terms of fragmentation densities 
and coefficient functions to $O(\alpha_s^2)$ in the unpolarized and the polarized case.
In section~3 we derive the Mellin moments for the coefficient functions and discuss their
structure w.r.t. the set of basic functions emerging. It turns out that this is a subset 
of the functions needed in the  space--like and related cases \cite{JBSMx,Blumlein:2005im}.
The analytic continuations for these functions to complex values of the Mellin variable $N$
was given before in \cite{Blumlein:2000hw}. Section 4 contains the conclusions.  
In the Appendices~A and B the explicit representations of the Wilson coefficients in the
unpolarized and polarized case in Mellin space are presented.
%%%%%%%%%%%%%%%%%%%%%%%%%%%%%%%%%%%%%%%%%%%%%%%%%%%%%%%%%%%%%%%%%%%%%%%%%%%%%%%%%%%%%%%%%%%%%%%%   
\section{Coefficient Functions}
%%%%%%%%%%%%%%%%%%%%%%%%%%%%%%%%%%%%%%%%%%%%%%%%%%%%%%%%%%%%%%%%%%%%%%%%%%%%%%%%%%%%%%%%%%%%%%%%   

\vspace{1mm}\noindent
Let us consider the differential scattering cross section for $e^+e^-$--annihilation into a 
hadron $H$ and the hadronic remainder part $X$
%-----------------------------------------------------------------------------------------------
\begin{eqnarray}
e^+ + e^- \rightarrow \gamma, Z \rightarrow H + `X'~.
\end{eqnarray}
%-----------------------------------------------------------------------------------------------
This reaction is either studied for unpolarized or polarized leptons. If in the latter case the
polarization of the produced hadron $H$ is measured, one can form the polarization asymmetry.
In the following we will consider both the case of unpolarized scattering and the 
final state polarization asymmetry in the polarized case.
%%%%%%%%%%%%%%%%%%%%%%%%%%%%%%%%%%%%%%%%%%%%%%%%%%%%%%%%%%%%%%%%%%%%%%%%%%%%%%%%%%%%%%%%%%%%%%%%   
\subsection{Unpolarized Case}
%%%%%%%%%%%%%%%%%%%%%%%%%%%%%%%%%%%%%%%%%%%%%%%%%%%%%%%%%%%%%%%%%%%%%%%%%%%%%%%%%%%%%%%%%%%%%%%%   

\vspace{1mm}\noindent
%-----------------------------------------------------------------------------------------------
\begin{eqnarray}
\label{disc} 
\frac{d^2 \sigma^H}{dx d \cos\theta} = \frac{3}{8} \left(1 + \cos^2 \theta\right)
                                       \frac{d \sigma_T^H}{dx}
                                     + \frac{3}{4} \sin^2 \theta 
                                       \frac{d \sigma_L^H}{dx}
                                     + \frac{3}{4} \cos \theta 
                                       \frac{d \sigma_L^A}{dx}~.
\end{eqnarray}
%-----------------------------------------------------------------------------------------------
Here $x = 2 p.q/Q^2$ denotes the scaling variable with $Q^2 = q^2 > 0$ and $\theta$ is scattering angel 
of the produced hadron. $p$ denotes the momentum of the 
produced hadron and $q$ is the virtuality of the exchanged $\gamma (Z)$. $x$ denotes the 
fraction of beam energy carried away by the produced hadron $H$. The differential scattering 
cross section (\ref{disc}) consists of three contributions $\sigma_{T,L,A}^H$ corresponding to 
the transverse $(T)$, longitudinal $(L)$ and the asymmetric $(A)$ part due to 
$\gamma-Z$ interference and pure $Z$--boson exchange.
The individual contributions to the differential cross sections are given by
%-----------------------------------------------------------------------------------------------
\begin{eqnarray}
\label{sigAd}
\frac{d \sigma^H_{k}}{dx} &=& \sigma_{\rm tot}^{(0)}(Q^2)
\Biggl[
D_{\rm S}^H(x,M^2) \otimes \C_{k,q}^{\rm S}\left(x,\frac{Q^2}{M^2}\right) +   
D_g^H(x,M^2) \otimes \C_{k,g}^{\rm S}\left(x,\frac{Q^2}{M^2}\right)\Biggr]
\nonumber\\
& & \qquad \qquad + \sum_{l=1}^{N_f} \sigma_l^{(0)}(Q^2) D_{{\rm NS},l}^H(x,M^2) \otimes 
\C_{k,q}^{\rm NS}\left(x,\frac{Q^2}{M^2}\right)~,
\end{eqnarray}
%-----------------------------------------------------------------------------------------------
with $k= T,L$. The asymmetric cross section is given by
%-----------------------------------------------------------------------------------------------
\begin{eqnarray}
\label{sigTLd}
\frac{d \sigma^H_{A}}{dx} &=& 
\sum_{l=1}^{N_f} A_l^{(0)}(Q^2) D_{{\rm A},l}^H(x,M^2) \otimes 
\C_{{\rm A},q}^{\rm NS}\left(x,\frac{Q^2}{M^2}\right)~.
\end{eqnarray}
%-----------------------------------------------------------------------------------------------
$N_f$ denotes the number of flavors. The Mellin convolution $\otimes$ is defined by
%-----------------------------------------------------------------------------------------------
\begin{eqnarray}
\label{mel}
\left[A \otimes B\right](x) = \int_0^1 dx_1 \int_0^1 dx_2 \delta(x - x_1 x_2) A(x_1) B(x_2)~.
\end{eqnarray}
%-----------------------------------------------------------------------------------------------
In (\ref{sigAd}, \ref{sigTLd}) 
the flavor singlet and non--singlet combinations of the 
non--perturbative fragmentation densities $D_i(z,M^2)$ contribute, which are given by
%-----------------------------------------------------------------------------------------------
\begin{eqnarray}
\label{frag}
D_{\rm S}^H(z,M^2) &=& \frac{1}{N_f} \sum_{l=1}^{N_f} \left[D_l^H(z,M^2) + 
D_{\overline{l}}(z,M^2) \right],\\
D_{{\rm NS},l}^H(z,M^2) &=& D_l^H(z,M^2) + 
D_{\overline{l}}(z,M^2) - D_{\rm S}(z,M^2),\\ 
D_{{\rm A},l}^H(z,M^2) &=& D_l^H(z,M^2) - 
D_{\overline{l}}(z,M^2)~, 
\end{eqnarray}
%-----------------------------------------------------------------------------------------------
with $M^2$ the factorization scale. The electro--weak pointlike cross sections are
%-----------------------------------------------------------------------------------------------
\begin{eqnarray}
\label{pointS}
\sigma_l^{(0)}(Q^2) &=& \frac{4 \pi \alpha^2}{3 Q^2} N_C \Biggl[e_L^2 e_l^2
+ \frac{2 Q^2(Q^2 - M_Z^2)}{|Z(Q^2)|^2} e_L e_l C_{V,L} C_{V,l} \nonumber\\
& & \qquad \qquad + \frac{(Q^2)^2}{|Z(Q^2)|^2} (C^2_{V,L}+C^2_{A,L}) (C_{V,l}^2 +C_{A,l}^2) 
\Biggr],\\
\sigma_{\rm tot}^{(0)}(Q^2) &=& 
\sum_{l=1}{N_f} \sigma_l^{(0)}(Q^2), 
\end{eqnarray}
%-----------------------------------------------------------------------------------------------
with
%-----------------------------------------------------------------------------------------------
\begin{eqnarray}
Z(Q^2) = Q^2 - M^2_Z + i M_Z \Gamma_Z~,
\end{eqnarray}
%-----------------------------------------------------------------------------------------------
and $M_Z, \Gamma_Z$ denote the mass and width of the $Z$--boson. The electric charges are $e_L = 
-1, e_u = 2/3, e_d = -1/3$ and $C_{V,(A),L}, C_{V,(A),l}$ are the electro--weak couplings of the 
charged lepton and the quarks, respectively. $N_C = 3$ denotes the number of 
colors. Correspondingly the asymmetry factor reads 
%-----------------------------------------------------------------------------------------------
\begin{eqnarray}
\label{pointA}
A_l^{(0)} =  \frac{4 \pi \alpha^2}{3 Q^2} N_C \Biggl[
\frac{2 Q^2(Q^2 - M_Z^2)}{|Z(Q^2)|^2} e_L e_l C_{A,L} C_{A,l}+ 4 
\frac{(Q^2)^2}{|Z(Q^2)|^2} C_{A,L} C_{A,l} C_{V,L} C_{V,l} 
\Biggr]~. 
\end{eqnarray}
%-----------------------------------------------------------------------------------------------

The  non--singlet and singlet coefficient functions are denoted by 
$\C_{l,q(g)}^{\rm S}(x,Q^2/M^2)$ and 
$\C_{l,q{\rm (A)}}^{\rm NS}(x,Q^2/M^2)$. To $O(a_s^2)$ they are given by 
%-----------------------------------------------------------------------------------------------
\begin{eqnarray}
\C_{L,q}^{\rm NS} &=& a_s c_{L,q}^{(1)} + a_s^2 \Biggl\{\left[ \frac{1}{2} P_{qq}^{(0)} \otimes 
c_{L,q}^{(1)} - \beta_0 c_{L,q}^{(1)} \right] \ln\left(\frac{Q^2}{M^2}\right)
+c_{L,q}^{{\rm NS},(2),nid}
+c_{L,q}^{{\rm NS},(2),id}\Biggr\}~,\\
%---------------------
\C_{T,q}^{\rm NS} &=& 1 + a_s \left[\frac{1}{2} P_{qq}^{(0)} \ln\left(\frac{Q^2}{M^2}\right)
+ c_{T,q}^{(1)}\right]  + a_s^2 \Biggl\{\left[ \frac{1}{8} 
P_{qq}^{(0)} \otimes P_{qq}^{(0)} - \frac{1}{4} \beta_0 P_{qq}^{(0)}\right] 
\ln^2\left(\frac{Q^2}{M^2}\right) \nonumber\\
& & + \left[ 
\frac{1}{2} \left(P_{qq}^{(1),{\rm NS}} + P_{q \overline{q}}^{(1),{\rm NS}}\right) 
- \beta_0 c_{T,q}^{(1)} + \frac{1}{2} P_{qq}^{(0)} \otimes c_{T,q}^{(1)} \right] 
\ln\left(
\frac{Q^2}{M^2}\right)\nonumber\\ & &
+ c_{T,q}^{{\rm NS},(2),nid} + 
  c_{T,q}^{{\rm NS},(2),id} \Biggr\}~, 
\\
\C_{A,q}^{\rm NS} &=& \C_{A,q}^ {{\rm NS},nid(2)} - \C_{A,q}^ {{\rm NS},id(2)}~. 
%---------------------
\end{eqnarray}
and
%-----------------------------------------------------------------------------------------------
\begin{eqnarray}
\C_{k,F}^{\rm S}  &=& \C_{k,F}^{\rm NS} + \C_{k,F}^{\rm PS},~~~F=L,T\\
\C_{L,q}^{\rm PS} &=& N_f a_s^2 \left[\left\{\frac{1}{4} P_{qg}^{(0)} \otimes c_{L,g}^{(1)}\right\}
\ln\left(\frac{Q^2}{M^2}\right)
                      + c_{L,q}^{{\rm PS},(2)} \right] \\
\C_{T,q}^{\rm PS} &=& N_f a_s^2 \Biggl[\left\{\frac{1}{8} P_{gq}^{(0)} \otimes P_{qg}^{(0)}  
\right\} \ln^2\left(\frac{Q^2}{M^2}\right)
+\left\{\frac{1}{2} P_{qq}^{{\rm PS},(1)} + \frac{1}{4} P_{qg}^{(0)} \otimes c_{T,g}^{(1)} \right\}
\ln\left(\frac{Q^2}{M^2}\right)
\nonumber\\ & & 
                      + c_{T,q}^{{\rm PS},(2)} \Biggr] \\
\C_{L,g}^{\rm S} &=& a_s c_{L,g}^{(1)} + a_s^2 \left[\left\{
- \beta_0 c_{L,g}^{(1)} + \frac{1}{2} P_{gg}^{(0)} \otimes c_{L,g}^{(1)}
+ P_{gq}^{(0)} \otimes c_{L,q}^{(1)}\right\} 
\ln\left(\frac{Q^2}{M^2}\right) + c_{L,g}^{(2)}\right]\\
\C_{T,g}^{\rm S} &=& a_s \left[P_{gq}^{(0)} 
\ln\left(\frac{Q^2}{M^2}\right) 
+ c_{T,g}^{(0)} \right] \nonumber\\
& & + a_s^2 \Biggl[ 
\left\{\frac{1}{4} P_{gq}^{(0)} \otimes \left ( 
 P_{gg}^{(0)}  
+P_{qq}^{(0)} \right) - \frac{1}{2} \beta_0 P_{gq}^{(0)} \right\}  
\ln^2\left(\frac{Q^2}{M^2}\right) 
\nonumber\\
& & + 
\left\{ P_{gq}^{(1)}
- \beta_0 c_{T,g}^{(1)} +\frac{1}{2} P_{gg}^{(0)} \otimes c_{T,g}^{(1)} 
+ P_{gq}^{(0)} \otimes c_{T,q}^{(1)} \right\}  
\ln\left(\frac{Q^2}{M^2}\right) 
+c_{T,g}^{(2)} \Biggr]
\end{eqnarray}
%-----------------------------------------------------------------------------------------------
Here $a_s = \alpha_s/(4\pi) = g_s^2/(4\pi)^2$ and $g_s$ denotes the strong coupling constant.
$P_{jk}^{(l)}$ are the the $l+1$-st order QCD splitting functions and $\beta_0 = 11 C_A/3 - (4/3)
T_R N_f$ with $C_A = 3, T_R = 1/2$ for $SU(3)_c$. The superscripts $`id'$ and $`nid'$ refer to 
identical quark contributions resp. the remainder part.
The corresponding  Wilson coefficients to $O(\alpha_s^2)$ were calculated in 
Refs.~\cite{Rijken:1996ns,Rijken:1996np,Rijken:1996vr}.

%%%%%%%%%%%%%%%%%%%%%%%%%%%%%%%%%%%%%%%%%%%%%%%%%%%%%%%%%%%%%%%%%%%%%%%%%%%%%%%%%%%%%%%%%%%%%%%%   
\subsection{Polarized Case}
%%%%%%%%%%%%%%%%%%%%%%%%%%%%%%%%%%%%%%%%%%%%%%%%%%%%%%%%%%%%%%%%%%%%%%%%%%%%%%%%%%%%%%%%%%%%%%%%   

\vspace{1mm}\noindent
The polarization asymmetry for the fragmentation process was calculated in \cite{Rijken:1997rg}
to $O(a_s^2)$. The hadronic tensor of the process is obtained from the hadronic tensor
in polarized electro--weak deeply inelastic scattering \cite{BKBT} by crossing from $t$ to 
$s$--channel. In general five fragmentation functions $g_i^H(x,Q^2)|_{i=1}^5$ 
contribute to the scattering cross section. 
If we limit the analysis to the level of twist--2 fragmentation functions 
the cross section is determined by $g_{1,4,5}^H$. Here $g_{4,5}^H$ are non--singlet
fragmentation functions which contribute to the $\gamma-Z$ and $|Z|^2$--exchange terms.
The corresponding coefficient functions are the same as for the fragmentation functions $F_2$ 
and $F_1$, respectively, in the unpolarized case due to the structure of the hadronic tensor,
cf.~\cite{BKBT}. In the following, we will therefore consider the case of pure 
photon exchange only.
If the scattering cross section is integrated over the azimuthal angel $\phi$
of the produced hadron the differential cross section is given by
%-----------------------------------------------------------------------------------------------
\begin{eqnarray}
\label{dsc-pol}
\frac{d \sigma^{H(\downarrow)}(\downarrow)}{dx d\cos \theta} -
\frac{d \sigma^{H(\downarrow)}(\uparrow)}{dx d\cos \theta} 
= N_C \frac{\pi \alpha^2}{Q^2} \cos \theta ~~ g_1^H(x,Q^2)~.
\end{eqnarray}
%-----------------------------------------------------------------------------------------------
The fragmentation function $g_1^H$ is given in terms of polarized fragmentation densities 
$\Delta_i^{H,{\rm S(NS)}}$ and Wilson coefficients $\Delta \C_{1,i}^{\rm S(NS)}$,
%-----------------------------------------------------------------------------------------------
\begin{eqnarray}
\label{eq-g1}
g_1^H(x,Q^2) = \frac{1}{N_f} \sum_{k=1}^{N_f} e_k^2 \Biggl[
\Delta_q^{H,{\rm S}}(x,M^2) \otimes
\C_{1,q}^{\rm S}\left(x,\frac{Q^2}{M^2}\right) 
+ \Delta_g^{H,{\rm S}}(x,M^2) \otimes
  \C_{T,g}^{\rm S}\left(x,\frac{Q^2}{M^2}\right) 
\nonumber\\
+ N_f \Delta_q^{H,{\rm NS}}(x,M^2) \otimes
      \C_{1,q}^{\rm NS}\left(x,\frac{Q^2}{M^2}\right) \Biggr]~.
\end{eqnarray}
%-----------------------------------------------------------------------------------------------
The polarized fragmentation densities are 
%-----------------------------------------------------------------------------------------------
\begin{eqnarray}
\label{pol-fr1}
\Delta_a^H(z,M^2) = 
 D_{a\downarrow}^{H\downarrow}(z,M^2)
-D_{a\downarrow}^{H\uparrow}(z,M^2),~~~a = q, \overline{q}, g~.
\end{eqnarray}
%-----------------------------------------------------------------------------------------------
The respective singlet and non--singlet combinations are given by
%-----------------------------------------------------------------------------------------------
\begin{eqnarray}
\label{pol-fr2}
\Delta_a^{H,{\rm S}}(z,M^2) = 
\sum_{k=1}^{N_f} 
  \Delta_{k}^{H}(z,M^2)
+ \Delta_{\overline{k}}^{H}(z,M^2)~,\\
\Delta_a^{H,{\rm NS}}(z,M^2) = 
  \Delta_{k}^{H}(z,M^2)
+ \Delta_{\overline{k}}^{H}(z,M^2) -
   \frac{1}{N_f}
\Delta_{q}^{H,{\rm S}}(z,M^2)~.
\end{eqnarray}
%-----------------------------------------------------------------------------------------------
The polarized non--singlet and singlet coefficient functions  
$\Delta \C_{1,a}^{\rm NS}(x,Q^2/M^2)$
and \newline $\Delta \C_{1,q(g)}^{\rm S}(x,Q^2/M^2)$ to $O(a_s^2)$ are given by 
%-----------------------------------------------------------------------------------------------
\begin{eqnarray}
\Delta \C_{1,q}^{\rm NS} &=& 1 + a_s \left[\frac{1}{2} \Delta P_{qq}^{(0)} 
\ln\left(\frac{Q^2}{M^2}\right)
+ \Delta c_{1,q}^{(1)}\right]  
\nonumber\\ & &
+ a_s^2 \Biggl\{\left[ \frac{1}{8} 
\Delta P_{qq}^{(0)} \otimes \Delta P_{qq}^{(0)} - \frac{1}{4} \beta_0 \Delta P_{qq}^{(0)}\right] 
\ln^2\left(\frac{Q^2}{M^2}\right) \nonumber\\
& & + \left[ 
\frac{1}{2} \left(\Delta P_{qq}^{(1),{\rm NS}} + \Delta P_{q \overline{q}}^{(1),{\rm NS}}\right) 
- \beta_0 c_{T,q}^{(1)} + \frac{1}{2} \Delta P_{qq}^{(0)} \otimes \Delta c_{1,q}^{(1)} \right] 
\ln\left(
\frac{Q^2}{M^2}\right)\nonumber\\ & &
+ \Delta c_{1,q}^{{\rm NS},(2),nid} + 
  \Delta c_{T,q}^{{\rm NS},(2),id} \Biggr\}~, 
\\
%---------------------
\Delta \C_{1,q}^{\rm PS} &=&  N_f a_s^2 \Biggl\{\left[ \frac{1}{8} 
\Delta P_{gq}^{(0)} \otimes \Delta P_{qg}^{(0)}\right]   \ln^2\left(\frac{Q^2}{M^2}\right) 
+ 
\left[ 
\frac{1}{2} \Delta P_{qq}^{(1),{\rm PS}} 
+ \frac{1}{4} \Delta P_{qg}^{(0)} \otimes \Delta c_{1,g}^{(1)} \right] 
\ln\left(
\frac{Q^2}{M^2}\right)\nonumber\\ & &
+ \Delta c_{1,q}^{{\rm PS},(2)}  
 \Biggr\}~, 
\\
%---------------------
\Delta \C_{1,g}^{\rm S} &=&  a_s \left[ \Delta P_{gq}^{(0)} 
\ln\left(\frac{Q^2}{M^2}\right)
+ \Delta c_{1,g}^{(1)}\right]  
\nonumber\\ & &
+ a_s^2 \Biggl\{\left[ \frac{1}{4} 
\Delta P_{gq}^{(0)} \otimes \left(
\Delta P_{gg}^{(0)} 
+ \Delta P_{qq}^{(0)}\right) 
- \frac{1}{2} \beta_0 \Delta P_{gq}^{(0)}\right] 
\ln^2\left(\frac{Q^2}{M^2}\right) \nonumber\\
& & + \left[ 
\Delta P_{gq}^{(1)} 
- \beta_0 c_{1,g}^{(1)} 
+ \frac{1}{2} \Delta P_{gg}^{(0)} \otimes \Delta c_{1,g}^{(1)}  
+             \Delta P_{gq}^{(0)} \otimes \Delta c_{1,q}^{(1)} \right] 
\ln\left(
\frac{Q^2}{M^2}\right)
+ \Delta c_{1,g}^{(2)}  
   \Biggr\}~, 
\end{eqnarray}
%-----------------------------------------------------------------------------------------------
with
%-----------------------------------------------------------------------------------------------
\begin{eqnarray}
\Delta \C^{\rm S}_{1,q} =
\Delta \C^{\rm NS}_{1,q} +
\Delta \C^{\rm PS}_{1,q}~.
\end{eqnarray}
%-----------------------------------------------------------------------------------------------
Here $\Delta P_{i,j}^{(k)}$ denote the polarized splitting functions.
The non--singlet Wilson coefficients for the structure functions $g_{1,4,5}$
are the same as in the unpolarized case for the corresponding tensor structures, i.e.
%-----------------------------------------------------------------------------------------------
\begin{eqnarray}
\Delta \C^{\rm NS}_{1,q} &=& \C^{\rm NS}_{1,q}\\
\Delta \C^{\rm NS}_{4,q} &=& \C^{\rm NS}_{2,q}\\
\Delta \C^{\rm NS}_{5,q} &=& \C^{\rm NS}_{1,q}~.
\end{eqnarray}
%-----------------------------------------------------------------------------------------------
The structure functions $g_{4,5}$ are pure non--singlet.

%%%%%%%%%%%%%%%%%%%%%%%%%%%%%%%%%%%%%%%%%%%%%%%%%%%%%%%%%%%%%%%%%%%%%%%%%%
\section{Mellin Moments}
%%%%%%%%%%%%%%%%%%%%%%%%%%%%%%%%%%%%%%%%%%%%%%%%%%%%%%%%%%%%%%%%%%%%%%%%%%

\vspace{1mm}\noindent
The coefficient functions to $O(\alpha_s^2)$ in $x$--space can be expressed in terms
of Nielsen integrals 
%-------------------------------------------------------------------------------------
\begin{equation}
\label{eq3aa}
S_{n,p}(y)\!\!=\!\!{(-1)^{n+p-1} \over (n-1)!p!}
\int_0^1 {dz \over z}\! \ln^{n-1}(z)\!
\ln^p(1-zy),
\end{equation} 
%-------------------------------------------------------------------------------------
partly with $y$ being a rational function of 
$x$~\cite{Rijken:1996ns,Rijken:1996np,Rijken:1996vr,Rijken:1997rg}. Here a relatively large
number of functions contributes.  
The  {\sf weight w} of these functions is defined by
{\sf w = p + n}, where any power of a logarithm counts for {\sf w = 1} as is the case for
the denominators $1/x,~1/(1+x)$ and $1/(1-x)$. In a product of functions 
the weights of the factors add. The Nielsen integrals cover the polylogarithms 
$S_{n-1,1}(y)=\Li_n(y)$ and through $d\Li_2(y)/d\ln(y) = - \ln(1-y)$ the logarithm.

For massless calculations it is suitable to consider the
coefficient functions in Mellin space. The  Mellin--transform of a function 
$F(x)$ is defined by
%-------------------------------------------------------------------------------------
\begin{equation}
{\Mvec}\big[F\big](N)=\int_0^1 dx x^{N-1} F(x)~.
\end{equation}
%-------------------------------------------------------------------------------------
In particular, for the Mellin  convolution $\left[F_1 \otimes F_2\right] (x)$
of two functions $F_1(x), F_2(x)$ 
one obtains
%-------------------------------------------------------------------------------------
\begin{eqnarray} 
{\Mvec}\big[F_1 \otimes F_2\big](N) = {\Mvec}\big[F_1\big](N) 
\cdot
{\Mvec}\big[F_2\big](N)~. 
\end{eqnarray}
%-------------------------------------------------------------------------------------
In Mellin space fragmentation densities and the corresponding Wilson coefficients 
are connected multiplicatively. Mellin space representations have also the advantage
that the QCD evolution equations can be solved analytically \cite{Gluck:1991ee,Blumlein:1997em} 
which are numerically very precise and fast.

The Mellin transforms of single-scale quantities in QCD can be represented in terms of nested
multiple harmonic sums \cite{Blumlein:1998if,Vermaseren:1998uu} and rational functions of the Mellin 
variable $N$. 
In this way a standardization is obtained. Harmonic sums are
described recursively by
%-------------------------------------------------------------------------------------
\begin{equation}
S_{b,a_1, \ldots, a_k}(N) = \sum_{l=1}^N \frac{{{\rm sign}(b)}^l}{l^{|b|}} S_{a_1, \ldots, a_k}(l)~,
\end{equation}
%-------------------------------------------------------------------------------------
where $b, a_1, \ldots, a_k$ are positive or negative integers and $N$ is a positive integer.
$S_a(N)$ is called single harmonic sum.

The harmonic sums $S_{a_1, \ldots, a_k}(N)$ form a Hopf algebra \cite{Moch:2001zr,Blumlein:2003gb}. 
For each index set 
$\{a_1, \ldots, a_k\}$ one can determine the basis elements through which all harmonic sums belonging to this 
set can be expressed algebraically. The number of these basic sums is given by the number of Lyndon words of 
this set and can be calculated by Witt-formulae \cite{WITT,Blumlein:2003gb}. 

For the application of the Mellin-space representation to experimental data one needs to derive
analytic continuations of the harmonic sums from integer values of $N$, $N$ - even or odd -, to complex 
values. While in the case of single harmonic sums these representations are known, they are more difficult to 
derive for multiple harmonic sums \cite{Blumlein:2000hw,Blumlein:2005jg}. 
As finally the argument $N$ will be complex one may consider further relations between harmonic sums for
rational and real values of the argument $N$ \cite{STRUCT} beyond the algebraic relations. In this way 
the amount of basic sums is further reduced. In particular one may consider equivalence classes of functions, 
which are related by differentiation. If an analytic representation for a harmonic sum is found in 
the complex plane outside its singularities, any derivation is easily obtained. It is convenient to choose 
for the basic functions Mellin transforms of a suitable harmonic polylogarithm \cite{Remiddi:1999ew} or 
Nielsen integrals with argument $x$ weighted by $1/(1 \mp x)$.
 
In former analyses of the two--loop anomalous dimensions, the two--loop Wilson coefficients for unpolarized 
and polarized deeply inelastic structure 
functions \cite{JBSMx}, the unpolarized and polarized Drell-Yan process and scalar and pseudoscalar Higgs 
boson production \cite{Blumlein:2005im}, $O(\alpha_s^3)$ heavy flavor contributions to $F_L(x,Q^2)$ 
\cite{JB05x}
and the 
unpolarized $O(\alpha_s^3)$ anomalous dimensions \cite{Blumlein:2005jg}
 the following minimal sets basic 
functions 
w.r.t the equivalence classes defined above were found. We list the newly emerging basic functions. 

\vspace{2mm}\noindent
{i)~~{\rm 1~loop~anomalous~dimensions~and~coefficient~functions~:}\footnote{In case of functions
which diverge as $x \rightarrow 1$ the $(...)_+$ prescription is understood.} 
%----------------------------------------------------------------------------------------------------------
\begin{eqnarray}
\Mvec\left[\frac{1}{1-x}\right](N)
\end{eqnarray}
%----------------------------------------------------------------------------------------------------------
{ii)~~{\rm 2~loop~anomalous~dimensions~:} \nonumber\\
%----------------------------------------------------------------------------------------------------------
\begin{eqnarray}
A_3(N) = \Mvec\left[\frac{\Li_2(x)}{1+x}\right](N)
\end{eqnarray}
%----------------------------------------------------------------------------------------------------------
{iii)~~{\rm 2~loop~coefficient functions~:} \nonumber\\
%----------------------------------------------------------------------------------------------------------
\begin{eqnarray}
A_{18}(N) &=& \Mvec\left[\frac{\Li_2(x)}{1-x}\right](N)~~~~~~~~~~~~~
A_6(N) = \Mvec\left[\frac{\Li_3(x)}{1+x}\right](N)
\nonumber\\
A_{21}(N) &=& \Mvec\left[\frac{S_{1,2}(x)}{1 - x}\right](N)~~~~~~~~~~~~~
A_{8}(N) = \Mvec\left[\frac{S_{1,2}(x)}{1 + x}\right](N)~.
\end{eqnarray}
%---------------------------------------------------------------------------------------
In the case of the 3--loop anomalous dimensions further eight basic functions contribute 
\cite{Blumlein:2005jg}. The notation above follows Ref.~\cite{Blumlein:2000hw}. Since
%-------------------------------------------------------------------------
\begin{eqnarray}
\frac{\partial^k}{\partial N^k} \Mvec[f(x)](N) = \Mvec\left[\ln^k(x) 
f(x)\right](N)~,
\end{eqnarray}
%-------------------------------------------------------------------------
one obtains
%-------------------------------------------------------------------------
\begin{eqnarray}
\label{der1}
A_5(N) &=& \frac{\partial}{\partial N} A_3(N) \\
\label{der2}
A_{22}(N) &=& \frac{\partial}{\partial N} A_{18}(N)~. 
\end{eqnarray}
%-------------------------------------------------------------------------
$A_{5}(N)$ and $A_{22}(N)$ are represented by $A_{3}(N)$ and $A_{18}(N)$ 
in the sense of equivalence classes as outlined above. 

The Mellin transforms of the unpolarized and polarized time--like coefficient functions to $O(\alpha_s^2)$ 
are given in Appendix A and B. Here we discuss their principal structure. Unlike the case
for the $O(\alpha_s^2)$ coefficient functions for deeply inelastic scattering, the process related through 
crossing from the time--like case considered in this paper to the space like-region
\cite{BNR}, besides the single harmonic sums only three other basic function contribute.
For the deep-inelastic Wilson coefficients \cite{JBSMx} and the polarized and unpolarized Drell-Yan 
process and 
scalar and pseudoscalar Higgs-boson production cross sections in the heavy mass limit \cite{Blumlein:2005im} five 
additional functions occurred. As the corrections to the latter hadronic processes are initial-state 
corrections they belong to the same class as space--like processes. 
Furthermore it is interesting to see, whether any of the harmonic sums contributing contains an index $\{-1\}$
since those sums were generally absent in all cases analyzed before.
Alternating single harmonic sums are given by 
%-------------------------------------------------------------------------
\begin{eqnarray}
\label{al1}
S_{-1}(N) &=& (-1)^N \beta(N+1) - \ln(2) \\
\label{al2}
S_{-k}(N) &=& (-1)^{(N+1+k)} \frac{1}{(k-1)!} \beta^{(k-1)}(N+1) 
- \left(1- \frac{1}{2^{k-1}}\right) \zeta(k),~~~ k > 1~,
\end{eqnarray}
%-------------------------------------------------------------------------
with
%-------------------------------------------------------------------------
\begin{eqnarray}
\beta(x) = \frac{1}{2} \left[\psi\left(\frac{x+1}{2}\right) 
- \psi\left(\frac{x}{2}\right)\right]
\end{eqnarray}
%-------------------------------------------------------------------------
and $\psi(x) = d \ln(\Gamma(x))/dx$.
In Appendix A and B  the $\beta-$function indeed occurs. However, it emerges always 
together with the function $A_3(N)~:$
%-------------------------------------------------------------------------
\begin{eqnarray}
A_3(N) &=& \Mvec\left[ \frac{\Li_2(x)}{1+x}\right](N) 
%\nonumber\\
%&=& 
=(-1)^{N+1}
\left[S_{-2,1}(N) - \zeta(2) S_{-1}(N)
+ \frac{5}{8} \zeta(3) - \zeta(2) \ln(2)\right]~,
\end{eqnarray}
%-------------------------------------------------------------------------
in such a way, that it cancels. Therefore, all contributing harmonic sums 
do not contain the index $\{-1\}$ also in the present cases.

The analytic continuations of the Wilson coefficients from integer values to complex values of 
$N$ is now easily performed. The alternating single sums are expressed in terms of Eqs.~(\ref{al1},\ref{al2}).
The non--alternating single sums obey the representation 
%-------------------------------------------------------------------------
\begin{eqnarray}
\label{al3}
S_{1}(N) &=& \psi(N+1) +\gamma_E \\
\label{al4}
S_{k}(N) &=& (-1)^{(1+k)} \frac{1}{(k-1)!} \psi^{(k-1)}(N+1) + \zeta(k)~. 
\end{eqnarray}
%-------------------------------------------------------------------------
The analytic continuations for the remaining functions  $A_3(N), A_{18}(N)$ and
$A_{21}(N)$ were performed in \cite{Blumlein:2000hw} and $A_5(N)$ and $A_{22}(N)$
are obtained as derivatives (\ref{der1},\ref{der2}).

In the case of the longitudinal fragmentation functions a transformation to Mellin space 
was given  in \cite{Albino:2005me} recently using the transformation tables of one of the authors
\cite{Blumlein:1998if}, see also \cite{Blumlein:1997vf}. In Appendix A and B we present the
Mellin transform in such a way, that factors of $(-1)^N$ still present 
in \cite{Albino:2005me}, Eq.~(A5), do not occur, and use further compactifications compared 
to earlier results.
As well known, the analytic continuation of the Mellin moments is carried out 
either from the even or odd integer values of $N$ determined by the  crossing 
relations $q \rightarrow -q, p \rightarrow p$ of the  respective quantity, cf.~\cite{BKBT}.
Thus, factors of $(-1)^N$ are not present in the analytic continuation. 
 
A {\tt FORTRAN}-programme of the 2--loop Wilson coefficients for the representation of the 
fragmentation functions for complex values of the Mellin variable suitable to extend analyses 
of the experimental fragmentation data \cite{Albino:2005me,EXP} can be obtained by the authors.

%%%%%%%%%%%%%%%%%%%%%%%%%%%%%%%%%%%%%%%%%%%%%%%%%%%%%%%%%%%%%%%%%%%%%%%%%%%%%%%%%%%%%%%%%%%%% 
\section{Conclusion}
%%%%%%%%%%%%%%%%%%%%%%%%%%%%%%%%%%%%%%%%%%%%%%%%%%%%%%%%%%%%%%%%%%%%%%%%%%%%%%%%%%%%%%%%%%%%% 

\vspace{1mm}\noindent
The mathematical structure of the 2--loop Wilson coefficients for the polarized 
and unpolarized time--like fragmentation functions simplifies considerably in Mellin 
space expressing the result in terms of nested harmonic sums. Using algebraic and 
structural identities for the harmonic sums and their continuations to rational, real and
finally complex values of the Mellin variable. The physical evolution kernels for
deeply inelastic scattering and parton fragmentation into hadrons are related by crossing
from $t-$ to $s-$channel order by order in perturbation theory. Since these quantities
are rational functions of the respective Wilson coefficients and anomalous dimensions
in Mellin space one expects a familiarity of the mathematical structure of the
space- and time--like Wilson coefficients. While in the space--like case five non-trivial
basic functions emerge besides the single harmonic sums, in the time--like case only three of 
them contribute to $O(\alpha_s^2)$. In $x$--space many more functions are required to 
express the Wilson coefficients. As observed for various other processes, harmonic sums
with index $\{-1\}$ do not contribute. The Mellin-space expressions allow to use the
time--like Wilson coefficients in fast evolution codes for the experimental analysis 
of fragmentation function data.

\vspace{2mm}\noindent
{\sf Note added.}~~The time-like Wilson coefficients in the unpolarized case have
been newly calculated in \cite{SMAM} very recently.}

\vspace{3mm}\noindent
{\bf Acknowledgment.}\\
We would like to thank W.L. van Neerven for discussions and code-comparison and  S. Moch 
for discussions. V.R. would like to thank DESY for their kind hospitality extended to him.  
This paper was supported in part by DFG Sonderforschungsbereich Transregio 9, Computergest\"utzte 
Theoretische Physik.}

\newpage
%%%%%%%%%%%%%%%%%%%%%%%%%%%%%%%%%%%%%%%%%%%%%%%%%%%%%%%%%%%%%%%%%%%%%%%% 
\begin{appendix}
\section{Unpolarized Time--like Coefficient Functions}

\label{sec-A}
\renewcommand{\theequation}{\thesection.\arabic{equation}}
\setcounter{equation}{0}
%\section{A}

\vspace{1mm}\noindent
In this appendix, we present the Mellin moments of unpolarized time--like  coefficient functions to
$O(\alpha_s^2)$~:

%%%%%%%%%%%%%%%%%%%%%%%%%%%%%%%%%%%%%%%%%%%%%%%%%%%%%%%%%%%%%%%%%%%%%%%%%%%%%%%%%%%%%%%%%%%%%%%%%%%
\begin{equation}
\C_{T,L,a}^{F} = C_{T,L,a}^{F,(0)}+a_s~ C_{T,L,a}^{F,(1)}+a_s^2~ C_{T,L,a}^{F,(2)} 
\end{equation}
with $F = {\rm NS,PS,S}$.

\vspace{1mm}\noindent
The individual contributions are
%--------------------------------------------------------------------------------------------------
\begin{eqnarray}
C_{T,q}^{{\rm NS}(0)} &=& 1
\\
C_{T,q}^{{\rm NS}(1)} &=& C_F \Bigg[  
                      \left( -\frac{2}{(N+1)}
                             - \frac{2}{N}+3-4 S_1(N-1)  
                      \right)  
                      \log\left(\frac{Q^2}{\mu^2}\right ) 
                     -9 + \frac{4}{N^2} - \frac{3}{N+1}  
\nonumber\\
              & & 
                       + \frac{3}{N} + \frac{4}{(N+1)^2}
                       + 2 S_1^2(N-1) + 2 \frac{ S_1 ( N+1 ) }{(N+1)} 
                       + 2 \frac { S_1( N ) }{N}  +3 S_1(N-1)            
\nonumber\\
              & & 
                       +10  S_2(N-1) \Bigg] 
\\
%-------------------------------------------------------------------------------------------------------
C_{T,g}^{{\rm S}(1)} &=& C_F \Bigg[
 \Bigg( \,{4 \over (N+1)}-\,{8 \over N}+\,{8 \over (N-1)} \Bigg) 
{ \log\left({Q^2 \over \mu^2}\right )}+\,{16 \over N^2}
- \,{ 16\over (-1+N)^2}+\,{8 \over N}
\nonumber\\&&
-\,{8 \over (N+1)^2}
-8\, {\frac {{ S_1} ( -1+N ) }{
(-1+N)}}-4\,{\frac {{ S_1} ( N+1 ) }{(N+1)}}+8\,
{\frac {{ S_1} ( N) }{N}}-\,{8 \over (-1+N)} \Bigg]
%\end{eqnarray}
\\
%{ c1qCfmCaCf}
%\begin{eqnarray}
\label{eqA5}
C_{T,q}^{{\rm NS}(2)}&=&C_F \Bigg(C_F-{C_A \over 2} \Bigg) \Bigg[
\,{36 \over N^4}
+\,{24 \over N^3}
-\,{36 \over (N+1)^4}
+\,{16 \over (N+1)^3}
+\,{82 \over 5 (N+1)^2}
\nonumber\\ &&
-\,{46 \over 5(N+1)}
-\,{16 \over (N+2)^3}
-\,{24 \over 5(N+2)^2}
+\,{24 \over 5 (N+2)}
+\,{24 \over 5(N+3)^3}
\nonumber\\ &&
-\,{24 \over 5(N-1)}
-\,{24 \over 5(N-1)^2}
-16\,{ \beta^{(1)}} ( N ) 
-{\frac {28}{3}}\,{ \beta^{(3)}} ( N ) 
-48\,{ A_5} ( N-1 ) 
\nonumber\\ &&
+16\,{\frac {{ \zeta_2} }{{(N+1)}^{2}}}
+4\,{\frac {{ \zeta_2}}{(N+1)}}
+8\,{\frac {{ \zeta_2}}{(N+2)}}
-{ \frac {12}{5}}\,{\frac {{ \zeta_2}}{(N+3)}}
-8\,{\frac {{ \zeta_2}}{(N-1)}}
\nonumber\\ &&
+{\frac { 12}{5}}\,{\frac {{ \zeta_2}}{(N-2)}}
-48\,{ S_2} ( N-1 ) { \beta^{(1)}} ( N ) 
+{\frac {24}{5}}\,{\frac {{ \beta^{(1)}} ( N-1 ) }{(N-2)}}
-16 \,{\frac {{ \beta^{(1)}} ( N+3 ) }{(N+2)}}
\nonumber\\ &&
+{\frac {24}{5}}\,{\frac {{ \beta^{(1)}} ( 4+N ) }{(N+3)}}
+72\,{ \beta^{(1)}} ( N ) { \zeta_2}
-16\,{\frac { { \beta^{(1)}} ( N ) }{(N-1)}}
-32\,{ \beta^{(1)}} ( N ) { S_1^2} ( N-1 ) 
\nonumber\\ &&
+32\,{\frac {{ \beta^{(2)}} ( N+2 ) }{(N+1)}}
+32\,{\frac {{ A_3} ( N+1 ) }{(N+1)}}
-32\,{\frac {{ \beta} ( N+2 ) { \zeta_2}}{(N+1)}}
+ \Bigg( 32\,{ S_1} ( N-1 ) { \beta^{(1)}} ( N ) 
\nonumber\\
%\end{eqnarray}\begin{eqnarray}
% 
&& -8 \,{ \beta^{(2)}} ( N ) 
+\,{8 \over (N+1)^3}
-\,{8 \over (N+1)^2}
+16\,{\frac {{ \beta^{(1)}} ( N+2 ) }{(N+1)}}
+32\,{ \beta} ( N ) { \zeta_2}
\nonumber\\ &&
-\,{16 \over (N+1)}
-32\,{ A_3} ( N-1 ) 
+{\frac {16-16\,{ \beta^{(1)}} ( N+1 ) }{ N}}
-\,{8 \over N^2}
-\,{8 \over N^3} \Bigg)  \log\left({Q^2 \over \mu^2}\right )
\nonumber\\ &&
+ \Bigg( -32\,{ \beta} ( N ) { \zeta_2}
+56\,{ \beta^{(2)}} ( N ) 
+32\,{ A_3} ( N-1 ) \Bigg) { S_1} ( N-1 ) 
+ \Bigg( -4\,{ \zeta_2}
-24\,{ \beta^{(2)}} ( N+1 ) 
\nonumber\\ &&
+{\frac {46}{5}} \Bigg) {1 \over N}
+ \Bigg( \,{8 \over (N+1)^2}
-\,{8 \over (N+1)^3}
-48 \,{\frac {{ \beta^{(1)}} ( N+2 ) }{(N+1)}}
+\,{16 \over (N+1)} \Bigg) { S_1} ( N+1 ) 
\nonumber\\ &&
+ \Bigg( -{\frac {158}{5}}
+16\,{ \beta^{(1)}} ( N+1 ) 
-8\,{ \zeta_2} \Bigg) {1 \over N^2}
+ \Bigg( {\frac {-16+16\,{ \beta^{(1)}} ( N+1 ) }{N} }
\nonumber\\ &&
+\,{8 \over N^2}+\,{8 \over N^3} \Bigg) { S_1} ( N ) \Bigg]
\nonumber\\ &&
%\end{eqnarray}
%%%%%%%%%%%%%%%%%%%%%%%%%%%%%%%%%%%%%%%%%%%%%%%%%%%%%%%%%%%%%%%%
%{ c1qnfCfTf}
%\begin{eqnarray}
+ N_F C_F T_F \Bigg[
 \Bigg( -{8 \over 3}\,{ S_1} ( N-1 ) 
-\,{4 \over 3 (N+1)}
-\,{4 \over 3 N}+2 \Bigg) \log^2\left({Q^2 \over \mu^2}\right )
+ \Bigg( {8 \over 3}\,{\frac {{ S_1} ( N+1 ) }{(N+1)}}
\nonumber\\ &&
+\,{28 \over 9 N}
+{8 \over 3}\,{\frac {{ S_1} ( N ) }{N}}
+8\,{ S_2} ( N-1) 
+\,{8 \over 3 (N+1)^2}
+{\frac {116}{9}}\,{ S_1} ( N-1 ) 
+\,{52 \over 9 (N+1)}
\nonumber\\ &&
+{8 \over 3}\,{ S_1^2} ( N-1 ) 
+\,{8 \over 3 N^2}
-{\frac {38}{3}}\Bigg) \log\left({Q^2 \over \mu^2}\right )
+ \Bigg( -{\frac {494}{27}}
+{16 \over 3}\,{ \zeta_2}
-{8 \over 3}\,{ S_2} ( N -1 )  \Bigg) { S_1} ( N-1 ) 
\nonumber\\ &&
+ \Bigg( -\,{8 \over 3 (N+1)^2}
-\,{52 \over 9 (N+1)} \Bigg) { S_1} ( N+1 ) 
+ \Bigg( -\,{8 \over 3 N^2}
-\,{28 \over 9 N} \Bigg) { S_1} ( N ) 
\nonumber\\ &&
-{\frac {58}{9}}\,{ S_1^2} ( N-1 ) 
-{4 \over 3}\,{\frac {{ S_1^2} ( N+1 ) }{(N+1)}}
-{4 \over 3}\,{\frac { { S_1^2} ( N ) }{N}}
-{\frac {8}{9}}\,{ S_1^3} ( N-1 ) 
-{\frac {250}{9}}\,{ S_2} ( N-1 ) 
\nonumber\\ &&
-{4 \over 3}\,{\frac {{ S_2} ( N+1 ) }{(N+1)}}
-{4 \over 3}\,{\frac {{ S_2} ( N ) }{N}}
-{\frac {40}{9}}\,{ S_3} ( N-1) 
-4\,{ \zeta_2}
+{16 \over 3}\,{ \zeta_3}
+{\frac {457}{18}}
-\,{34 \over 27 (N+1)}
\nonumber\\ &&
+{8 \over 3}\,{\frac {{ \zeta_2}}{(N+1)}}
-\,{4 \over 3 (N+1)^3}
-\,{12 \over (N+1)^2}
+{16 \over 3}\,{ A_{18}} ( N-1 ) 
+ \Bigg( {8 \over 3}\,{ \zeta_2}
-{\frac {118}{27}} \Bigg) {1 \over N} 
\nonumber\\ &&
-\,{20 \over 3 N^2}
-\,{4 \over 3 N^3} 
\Bigg] \nonumber\\
%\end{eqnarray}
%%%%%%%%%%%%%%%%%%%%%%%%%%%%%%%%%%%%%%%%%%%%%%%%%%%%%%%%%%%%%%%%%%%%%
%{ c1qCaCf}
%\begin{eqnarray}
&& +C_A C_F \Bigg[
 \Bigg(  \Big( 8\,{ \zeta_2}
-{\frac {367}{9}} \Big) { S_1} ( N-1 ) 
-{ \frac {22}{3}}\,{\frac {{ S_1} ( N+1 ) }{(N+1)}}
-{\frac {22}{3}}\,{\frac {{ S_1} ( N ) }{N}}
-{\frac {22}{3}}\,{ S_1^2} ( N-1 ) 
\nonumber\\ &&
-22\,{ S_2} ( N-1 ) 
-8\,{ S_3} ( N-1 ) 
+{\frac {215}{6}}
-\,{34 \over 3 (N+1)^2}
-\,{4 \over (N+1)^3}
+4\,{\frac {{ \zeta_2}}{(N+1)}}
\nonumber\\ &&
-4\,{ \zeta_3}
-\,{275 \over 9 (N+1)}
+ \Big( 4\,{ \zeta_2}
+{\frac {7}{9}} \Big) {1 \over N}
-\,{34 \over 3 N^2}
-\,{4 \over N^3} \Bigg) \log\left({Q^2 \over \mu^2}\right )
+3\,{ \zeta_2}
\nonumber\\ &&
+\,{18 \over N^4}
+\,{47 \over 3 N^3}
+\,{18 \over (N+1)^4}
+\,{47 \over 3 (N+1)^3}
+\,{184 \over 5(N+1)^2}
+\,{3587 \over 270 (N+1)}
\nonumber\\ &&
-\,{8 \over (N+2)^3}
+\,{12 \over 5 (N+2)^2}
-\,{12 \over 5 (N+2)}
-\,{12 \over 5 (N+3)^3}
+\,{12 \over 5 (N-1)}
\nonumber\\ &&
+\,{12 \over 5 (N-1)^2}
-8\,{\frac {{ \zeta_2}}{{(N+1)}^{2}}}
-{ \frac {22}{3}}\,{\frac {{ \zeta_2}}{(N+1)}}
+4\,{\frac {{ \zeta_2}}{(N+2)}}
+{6 \over 5}\,{ \frac {{ \zeta_2}}{(N+3)}}
\nonumber\\ &&
-4\,{\frac {{ \zeta_2}}{(N-1)}}
-{6 \over 5}\,{\frac {{ \zeta_2}}{ (N-2)}}
-{\frac {12}{5}}\,{\frac {{ \beta^{(1)}} ( N-1 ) }{(N-2)}}
-8\,{\frac {{ \beta^{(1)}} ( N+3 ) }{(N+2)}}
-{\frac {12}{5}}\,{\frac {{ \beta^{(1)}} ( 4+N) }{(N+3)}}
\nonumber\\ &&
-8\,{\frac {{ \beta^{(1)}} ( N ) }{(N-1)}}
+{\frac {118}{3}}\, { \zeta_3}
+{11 \over 3}\,{\frac {{ S_1^2} ( N+1 ) }{(N+1)}}
+{11 \over 3}\,{\frac {{ S_1^2} ( N ) }{N}}
+{\frac {110}{9}}\,{ S_3} ( N-1 ) 
\nonumber\\ &&
-{\frac {44} {3}}\,{ A_{18}} ( N-1 ) 
+{\frac {22}{9}}\,{ S_1^3} ( N-1 ) 
- 34\,{\frac {{ \zeta_3}}{(N+1)}}
+8\,{\frac {{ S_3} ( N+1 ) }{(N+1)}}
+16\, {\frac {{ \beta^{(1)}} ( N+2 ) }{{(N+1)}^{2}}}
\nonumber\\ &&
+4\,{\frac {{ \beta^{(1)}} ( N+2) }{(N+1)}}
-8\,{\frac {{ A_{18}} ( N+1 ) }{(N+1)}}
-12\, ( { \beta^{(1)}} ( N )  ) ^{2}
+24\,{ A_{21}} ( N-1 ) 
+44\,{ S_4} ( N-1 ) 
\nonumber\\ &&
+8\,{ S_2^2} ( N-1 ) 
-{\frac {97}{5}}\,{{ \zeta_2} }^{2}
-{\frac {5465}{72}}
+ \Bigg( -4\,{ S_2} ( N-1 ) 
+{\frac {367}{18}}\Bigg) { S_1^2} ( N-1 ) 
\nonumber\\ &&
+ \Bigg( 8\,{ S_3} ( N-1 ) 
+{\frac {22}{3}}\,{ S_2} ( N-1 ) 
-48\,{ \zeta_3}
+{\frac {3155}{54}}
-16\,{ A_{18}} ( N-1 ) 
\nonumber\\ &&
-{\frac {44}{3}}\,{ \zeta_2} \Bigg) { S_1} ( N-1) 
+ \Bigg( {\frac {1603}{18}}
-20\,{ \zeta_2} \Bigg) { S_2} ( N-1) 
+ \Bigg( \,{4 \over (N+1)^2}
\nonumber\\ &&
+\,{11 \over 3(N+1)} \Bigg) { S_2} ( N+1) 
+ \Bigg( -4\,{\frac {{ S_2} ( N+1 ) }{(N+1)}}
+\,{34 \over 3 (N+1)^2}
+\,{383 \over 9 (N+1)}
\nonumber\\ &&
-4\,{\frac {{ \zeta_2}}{(N+1)}} \Bigg) { S_1} ( N+1 ) 
+ \Bigg( \,{11 \over 3 N}
+\,{4 \over N^2} \Bigg) { S_2} ( N ) 
+ \Bigg( {\frac {2513}{270}}
-14\,{ \zeta_3}
-{\frac {22}{3}}\,{ \zeta_2}
\nonumber\\
% 
%\end{eqnarray}\begin{eqnarray}
&&
+4\,{ \beta^{(1)}} ( N+1 ) 
-8\,{ A_{18}} ( N )  \Bigg) {1 \over N}
+ \Bigg( {\frac {122}{15}}
-12\,{ \zeta_2}
+8\,{ \beta^{(1)}} ( N+1 ) \Bigg) {1 \over N^2}
+ \Bigg( \,{11 \over 3 N}
\nonumber\\ &&
-{11 \over 2}
+{\frac {22}{3}}\,{ S_1} ( N-1) 
+\,{11 \over 3 (N+1)} \Bigg) \log^2\left({Q^2 \over \mu^2}\right )
+ \Bigg( -4\,{\frac {{ S_2} ( N ) }{N}}
+ \Bigg( 4\,{ \zeta_2}-{\frac {97}{9}} \Bigg) {1 \over N}
\nonumber\\ &&
+\,{34 \over 3 N^2}\Bigg ) { S_1} ( N ) \Bigg]
%\end{eqnarray}
%%%%%%%%%%%%%%%%%%%%%%%%%%%%%%%%%%%%%%%%%%%%%%%%%%%%%%%%%%%%%%%%%%%%
\nonumber\\ &&
%\begin{eqnarray}
%{ c1qCf2}
+ C_F^2 \Bigg[
67\,{ \zeta_2}
-\,{102 \over N^4}
-\,{90 \over N^3}
-\,{102 \over (N+1)^4}
-\,{46 \over (N+1)^3}
-\,{248 \over 5 (N+1)^2}
\nonumber\\ &&
-\,{187 \over 10(N+1)}
+\,{16 \over (N+2)^3}
-\,{24 \over 5(N+2)^2}
+\,{24 \over 5(N+2)}
+\,{24 \over 5(N+3)^3}
\nonumber\\ &&
-\,{24 \over 5(N-1)}
-\,{24 \over 5(N-1)^2}
+ \Bigg( -8\,{\frac {{ \zeta_2}}{(N+1)}}
-\,{4 \over (N+1)^2}
+36\,{\frac {{ S_2} ( N+1 ) }{(N+1)}}
\nonumber\\ &&
-\,{16 \over (N+1)^3}
-\,{67 \over (N+1)} \Bigg) { S_1} ( N+1 ) 
+12\,{\frac {{ \zeta_2}}{{(N+1)}^{2}}}
-12\,{\frac {{ \zeta_2}}{(N+1)}}
-8\,{\frac {{ \zeta_2}}{(N+2)}}
\nonumber\\ &&
-{\frac {12}{ 5}}\,{\frac {{ \zeta_2}}{(N+3)}}
+8\,{\frac {{ \zeta_2}}{(N-1)}}
+{\frac {12}{5}}\,{ \frac {{ \zeta_2}}{(N-2)}}
+{\frac {24}{5}}\,{\frac {{ \beta^{(1)}} ( N-1 ) }{(N-2)}}
+16\,{\frac {{ \beta^{(1)}} ( N+3 ) }{(N+2)}}
\nonumber\\ &&
+{\frac {24}{5}}\,{\frac {{ \beta^{(1)}} ( 4+N ) }{(N+3)}}
+16\,{\frac {{ \beta^{(1)}} ( N ) }{(N-1) }}
-36\,{ \zeta_3}
-30\,{ S_3} ( N-1 ) 
+12\,{ A_{18}} ( N-1) 
\nonumber\\ &&
+6\,{ S_1^3} ( N-1 ) 
+20\,{\frac {{ \zeta_3}}{(N+1)}}
+4\,{\frac {{ S_3} ( N+1 ) }{(N+1)}}
-32\,{\frac {{ \beta^{(1)}} ( N+2 ) }{{(N+1)}^{2}}}
-8\,{\frac {{ \beta^{(1)}} ( N+2 ) }{(N+1)}}
\nonumber\\ &&
+8\,{\frac {{ A_{18}} ( N+1 ) }{(N+1)}}
+24\, ( { \beta^{(1)}} ( N )  ) ^{2}
-24 \,{ A_{21}} ( N-1 ) 
-92\,{ S_4} ( N-1 ) 
+54\,{ S_2^2} ( N-1 ) 
\nonumber\\ &&
+{\frac {38}{5}}\,{{ \zeta_2}}^{2}
+8\,{ A_{22}} ( N-1) 
+20\,{\frac {{ S_3} ( N ) }{N}}
+4\,{\frac {{ S_1^3} ( N+1) }{(N+1)}}
+4\,{\frac {{ S_1^3} ( N ) }{N}}
+2\,{ S_1^4} ( N-1) 
\nonumber\\ &&
+{\frac {331}{8}}
+ \Bigg( \,{8 \over (N+1)^2}
+\,{4 \over (N+1)} \Bigg) { S_1^2} ( N+1 ) 
+ \Bigg( -24\,{ \zeta_2}
-{\frac {51}{2}}
+42\,{ S_2} ( N-1) 
\nonumber\\ &&
+24\,{ S_3} ( N-1 ) 
+16\,{ A_{18}} ( N-1 ) \Bigg) { S_1} ( N-1 ) 
+ \Bigg( 36\,{\frac {{ S_2} ( N ) }{ N}}
+{\frac {-24\,{ \zeta_2} +29}{N}}
-\,{20 \over N^2}
\nonumber\\ &&
-\,{16 \over N^3} \Bigg) { S_1} ( N ) 
+ \Bigg( \,{8 \over N}
+\,{8 \over N^2} \Bigg) { S_1^2} ( N ) 
+ \Bigg( -16\,{ \zeta_2}
-{\frac {27}{2}}
+36\,{ S_2} ( N-1 )  \Bigg) { S_1^2} ( N-1 ) 
\nonumber\\
%\end{eqnarray}
%\begin{eqnarray}
&&
%\nonumber\\ &&
+ \Bigg( 8\,{ S_1^2} ( N-1 ) 
-\,{6 \over N^2}
+{9 \over 2}
+ 8\,{\frac {{ S_1} ( N+1 ) }{(N+1)}}
-12\,{ S_1} ( N-1 ) 
+8\,{ \frac {{ S_1} ( N ) }{N}}
-\,{10 \over N}
\nonumber\\ &&
-\,{6 \over (N+1)^2}
-\,{2 \over (N+1)}\Bigg) \log^2\left({Q^2 \over \mu^2}\right )
+ \Bigg( -{\frac {239}{2}}
+24\,{ \zeta_2} \Bigg) { S_2} ( N-1 ) 
+ \Bigg( \,{36 \over N}
\nonumber\\ &&
+\,{44 \over N^2} \Bigg) { S_2} ( N) 
+ \Bigg( \,{44 \over (N+1)^2}
-\,{24 \over (N+1)} \Bigg) { S_2} ( N+1) 
+ \Bigg( {\frac {187}{10}}
-20\,{ \zeta_3}
-12\,{ \zeta_2}
\nonumber\\ &&
+8\,{ A_{18}} ( N ) 
-8\,{ \beta^{(1)}} ( N+1 )  \Bigg) {1 \over N}
+ \Bigg( -{\frac {218}{5}}
+20\,{ \zeta_2}
-16\,{ \beta^{(1)}} ( N+1 )  \Bigg) {1 \over N^2}
\nonumber\\ &&
+ \Bigg( \Big( -56\,{ S_2} ( N-1 ) 
+45
+16\,{ \zeta_2} \Big) { S_1} ( N -1 ) 
+ \Bigg( \,{4 \over (N+1)}
-\,{12 \over (N+1)^2} \Bigg) { S_1} ( N+1) 
\nonumber\\ &&
+ \Bigg( -\,{12 \over N^2}
-\,{4 \over N} \Bigg) { S_1} ( N ) 
-6\,{ S_1^2} ( N-1 ) 
-12\,{\frac {{ S_1^2} ( N+1 ) }{(N+1)}}
-12\,{ \frac {{ S_1^2} ( N ) }{N}}
-8\,{ S_1^3} ( N-1 ) 
\nonumber\\ &&
+42\,{ S_2} ( N-1 ) 
-28\,{\frac {{ S_2} ( N+1 ) }{(N+1)}}
-28\,{\frac {{ S_2} ( N ) }{N}}
+16\,{ S_3} ( N-1 ) 
-24\,{ \zeta_2}
+8\,{ \zeta_3}
\nonumber\\ &&
+\,{31 \over (N+1)}
+\,{44 \over (N+1)^3}
-{\frac {51}{2}}
+8\,{\frac {{ \zeta_2}}{(N+1)}}
+\,{20 \over (N+1)^2}
+{\frac {8\,{ \zeta_2} +5}{N}}
\nonumber\\ &&
+\,{52 \over N^2}
+\,{44 \over N^3}\Bigg)  \log\left({Q^2 \over \mu^2}\right ) \Bigg]
\\
%\end{eqnarray}
%%%%%%%%%%%%%%%%%%%%%%%%%%%%%%%%%%%%%%%%%%%%%%%%%%%%%%%%%%%%%%%%%
%\begin{eqnarray}
%{ c1qnfCfTfPS}
C_{T,q}^{{\rm PS}(2)}&=& N_F C_F T_F \Bigg[
-\,{88 \over N^4}
-\,{28 \over N^3}
-\,{88 \over (N+1)^4}
-\,{28 \over (N+1)^3}
+\,{208 \over 3 (N+1)^2}
\nonumber\\ &&
+\,{140 \over 3 (N+1)}
+\,{32 \over 3 (N+2)^3}
+\,{256 \over 9 (N+2)^2}
+\,{1024 \over 27 (N+2)}
-\,{160 \over 27 (N-1)}
\nonumber\\ &&
+\,{32 \over 3 (N-1)^2}
+16\,{\frac {{ \zeta_2}}{{(N+1)}^{2}}}
-8\,{\frac {{ \zeta_2}}{(N+1)}}
+16\,{\frac {{ \zeta_2}}{(N+2)}}
-16\,{\frac {{ \zeta_2}}{(N-1)}}
\nonumber\\ &&
+{\frac {32}{3 }}\,{\frac {{ \beta^{(1)}} ( N+3 ) }{(N+2)}}
+{\frac {32}{3}}\,{\frac {{ \beta^{(1)} } ( N ) }{(N-1)}}
+32\,{\frac {{ \beta^{(1)}} ( N+2 ) }{(N+1)}}
+\,{128 \over 3 (N-1)^3}
+{16 \over 3}\,{\frac {{ S_2} ( N-1 ) }{(N-1)}}
\nonumber\\ &&
-{16 \over 3} \,{\frac {{ S_2} ( N+2 ) }{(N+2)}}
+{16 \over 3}\,{\frac {{ S_1^2} ( N-1) }{(N-1)}}
-{16 \over 3}\,{\frac {{ S_1^2} ( N+2 ) }{(N+2)}}
+ \Bigg( -{ \frac {32}{3}}\,{\frac {{ S_1} ( N-1 ) }{(N-1)}}
\nonumber\\ &&
+ \Bigg( \,{16 \over (N+1)^2} 
+\,{8 \over (N+1)} \Bigg) { S_1} ( N+1 ) 
+ \Bigg( \,{16 \over N^2}
-\,{8 \over N}\Bigg) { S_1} ( N ) 
+\,{136 \over 3 (N+1)}
%\nonumber\\ &&
\nonumber\\
%\end{eqnarray}
%\begin{eqnarray}
&&
-\,{64 \over 3 (N-1)^2}
+\,{8 \over N^2}
-\,{184 \over 3 N}
-\,{16 \over 3 (N-1)}
+\, {64 \over 3 (N+2)}
\nonumber\\ &&
+\,{48 \over (N+1)^3}
+\,{48 \over N^3}
+{\frac {32}{3}}\,{\frac {{ S_1} ( N+ 2 ) }{(N+2)}}
+\,{40 \over (N+1)^2}
+\,{32 \over 3 (N+2)^2} \Bigg) \log\left({Q^2 \over \mu^2}\right )
\nonumber\\ &&
+ \Bigg( -\,{8 \over N^2}
+\,{4 \over N} \Bigg) { S_1^2} ( N ) 
+ \Bigg( \,{16 \over 3 (N-1)}
+\,{64 \over 3 (N-1)^2} \Bigg) { S_1} ( N-1 ) 
\nonumber\\ &&
+ \Bigg( - \,{48 \over (N+1)^3}
-\,{40 \over (N+1)^2}
-\,{136 \over 3 (N+1)} \Bigg) { S_1} ( N+1 ) 
+ \Bigg( -\,{48 \over N^3}
\nonumber\\ &&
+\,{184 \over 3 N}
-\,{8 \over N^2}\Bigg) { S_1} ( N ) 
+ \Bigg( -\,{64 \over 3 (N+2)}
-\,{32 \over 3 (N+2)^2} \Bigg) { S_1} ( N+2 ) 
+ \Bigg( -\,{8 \over N^2}
\nonumber\\ &&
+\,{4 \over N} \Bigg) { S_2} ( N ) 
+ \Bigg( -\,{16 \over 3 (N+2)}
+\,{16 \over 3 (N-1)}
- \,{8 \over (N+1)^2}
+\,{4 \over N}
-\,{4 \over (N+1)}
\nonumber\\ &&
-\,{8 \over N^2} \Bigg) \log^2\left({Q^2 \over \mu^2}\right )
+ \Bigg( -\,{4 \over (N+1)}
-\,{8 \over (N+1)^2} \Bigg) { S_2} ( N+1 ) 
+ \Bigg( -\,{4 \over (N+1)}
\nonumber\\ &&
-\,{8 \over (N+1)^2} \Bigg) { S_1^2} ( N+1 ) 
+ \Bigg( -{ \frac {236}{3}}
+8\,{ \zeta_2}+32\,{ \beta^{(1)}} ( N+1 )  \Bigg) {1 \over N}
\nonumber\\ &&
+ \Bigg( {\frac {400}{3}}+16\,{ \zeta_2} \Bigg) {1 \over N^2} \Bigg]
\\
%\end{eqnarray}
%%%%%%%%%%%%%%%%%%%%%%%%%%%%%%%%%%%%%%%%%%%%%%%%%%%%%%%%%%%%%%%
%{ c1gCaCf}
%\begin{eqnarray}
C_{T,g}^{{\rm S}(2)}&=&C_A C_F \Bigg[
\,{176 \over N^4}
-\,{32 \over N^3}
+\,{248 \over (N+1)^4}
+\,{4 \over (N+1)^3}
-\,{172 \over 3 (N+1)^2}
\nonumber\\ &&
-\,{106 \over (N+1)}
-\,{32 \over 3 (N+2)^3}
-\,{256 \over 9 (N+2)^2}
-\,{928 \over 27 (N+2)}
+\,{4438 \over 27 (N-1)}
\nonumber\\ &&
-\,{496 \over 3 (N-1)^2}
-64\,{\frac {{ \zeta_2}}{{(N+1)}^{2}}}
+20\,{\frac {{ \zeta_2}}{(N+1)}}
-16\,{ \frac {{ \zeta_2}}{(N+2)}}
+96\,{\frac {{ \zeta_2}}{(N-1)}}
\nonumber\\ &&
-{\frac {32}{3}}\,{\frac { { \beta^{(1)}} ( N+3 ) }{(N+2)}}
-{\frac {80}{3}}\,{\frac {{ \beta^{(1)}} ( N) }{(N-1)}}
-16\,{\frac {{ \beta^{(2)}} ( N+2 ) }{(N+1)}}
+32\,{\frac {{ A_3} ( N+1 ) }{(N+1)}}
\nonumber\\ &&
-32\,{\frac {{ \beta} ( N+2 ) { \zeta_2}}{(N+1)}}
+60\,{\frac {{ \zeta_3}}{(N+1)}}
-{\frac {200}{3}}\,{\frac {{ S_3} ( N+1 ) }{(N+1)}}
+32\,{\frac {{ \beta^{(1)}} ( N+2 ) }{{(N+1)}^{2}} }
-56\,{\frac {{ \beta^{(1)}} ( N+2 ) }{(N+1)}}
\nonumber\\
%\end{eqnarray}
%\begin{eqnarray}
&&
\nonumber\\ &&
-16\,{\frac {{ A_{18}} ( N +1 ) }{(N+1)}}
+{\frac {400}{3}}\,{\frac {{ S_3} ( N ) }{N}}
-{4 \over 3}\,{ \frac {{ S_1^3} ( N+1 ) }{(N+1)}}
+{8 \over 3}\,{\frac {{ S_1^3} ( N ) }{N}}
-\,{464 \over 3 (N-1)^3}
\nonumber\\ &&
+{16 \over 3}\,{\frac {{ S_2} ( N+2 ) }{(N+2) }}
+{16 \over 3}\,{\frac {{ S_1^2} ( N+2 ) }{(N+2)}}
+\,{320 \over (N-1)^4}
-96\,{\frac { { \zeta_2}}{{(N-1)}^{2}}}
+80\,{\frac {{ \zeta_3}}{(N-1)}}
\nonumber\\ &&
-{\frac {352}{3}}\,{\frac { { S_3} ( N-1 ) }{(N-1)}}
+64\,{\frac {{ \beta^{(1)}} ( N ) }{{(N-1) }^{2}}}
-40\,{\frac {{ \beta^{(2)}} ( N ) }{(N-1)}}
+32\,{\frac {{ A_3} ( N-1 ) }{(N-1)}}
-32\,{\frac {{ A_{18}} ( N-1 ) }{(N-1)}}
\nonumber\\ &&
-{8 \over 3} \,{\frac {{ S_1^3} ( N-1 ) }{(N-1)}}
-32\,{\frac {{ \beta} ( N) { \zeta_2}}{(N-1)}}
+ \Bigg( -\,{96 \over N^3}
-\,{64 \over N^2}
-\,{112 \over (N+1)^3}
- \,{40 \over (N+1)^2}
\nonumber\\ &&
-\,{100 \over 3 (N+1)}
-\,{32 \over 3(N+2)^2}
-\,{32 \over 3 (N+2)}
+\,{188 \over 3 (N-1)}
+\,{400 \over 3 (N-1)^2}
\nonumber\\ &&
+8\,{ \frac {{ \zeta_2}}{(N+1)}}
+16\,{\frac {{ \zeta_2}}{(N-1)}}
-32\,{\frac {{ \beta^{(1)}} ( N ) }{(N-1)}}
+40\,{\frac {{ S_2} ( N+1 ) }{(N+1)}}
-80\,{ \frac {{ S_2} ( N ) }{N}}
+8\,{\frac {{ S_1^2} ( N+1 ) }{(N+1) }}
\nonumber\\ &&
-16\,{\frac {{ S_1^2} ( N ) }{N}}
-16\,{\frac {{ \beta^{(1)}} ( N+2) }{(N+1)}}
-\,{128 \over (N-1)^3}
+{\frac {344}{3}}\,{\frac {{ S_1} ( N-1) }{(N-1)}}
-{\frac {32}{3}}\,{\frac {{ S_1} ( N+2 ) }{(N+2)}}
\nonumber\\ &&
+80\, {\frac {{ S_2} ( N-1 ) }{(N-1)}}
+16\,{\frac {{ S_1^2} ( N-1) }{(N-1)}}
+ \Bigg( -{\frac {32}{3}}
-16\,{ \zeta_2}
-32\,{ \beta^{(1)}} ( N+1)  \Bigg) {1 \over N}
\nonumber\\ &&
+ \Bigg( -\,{16 \over (N+1)^2}
+8\,{1 \over (N+1)} \Bigg) { S_1} ( N+1 ) 
+ \Bigg( -\,{64 \over N^2}
-96\,{1 \over N} \Bigg) { S_1} ( N)  \Bigg) \log\left({Q^2 \over \mu^2}\right )
\nonumber\\ &&
+ \Bigg( 64\,{\frac {{ \zeta_2}}{(N-1)}}
-72\,{\frac {{ S_2} ( N-1 ) }{(N-1)}}
-\,{496 \over 3 (N-1)^2}
+\,{96 \over (N-1)^3}
\nonumber\\ &&
-\,{356 \over 3 (N-1)} \Bigg) { S_1} ( N-1 ) 
+ \Bigg( \,{120 \over N^2}
+\,{112 \over N} \Bigg) { S_2} ( N ) 
+ \Bigg( -16\,{\frac {{ S_1} ( N-1 ) }{(N-1)}}
\nonumber\\ &&
-8\,{\frac {{ S_1} ( N+1 ) }{(N+1)}}
+{16 \over 3 (N+2)}
+\,{16 \over (N-1)^2}
+\,{32 \over N}
-\,{124 \over 3 (N-1)}
+16\,{\frac {{ S_1} ( N ) }{N}}
\nonumber\\ &&
+\,{16 \over (N+1)^2}
+\,{4 \over (N+1)}
+\,{16 \over N^2} \Bigg) \log^2\left({Q^2 \over \mu^2}\right )
+ \Bigg( -36\,{\frac {{ S_2} ( N+1 ) }{(N+1)}}
-16\,{\frac { { \beta^{(1)}} ( N+2 ) }{(N+1)}}
\nonumber\\ &&
+\,{16 \over (N+1)^2}
+\,{4 \over 3 (N+1)}
+\,{96 \over (N+1)^3}
+40\,{\frac {{ \zeta_2}}{(N+1)}} \Bigg) { S_1} ( N+1 ) 
+ \Bigg( 72 \,{\frac {{ S_2} ( N ) }{N}}
\nonumber\\
%\end{eqnarray}\begin{eqnarray}
&&
%\nonumber\\ &&
+ \Bigg( {\frac {236}{3}}
-32\,{ \beta^{(1)}} ( N+1 ) 
-80\,{ \zeta_2} \Bigg) {1 \over N}
+\,{96 \over N^2}
+\,{128 \over N^3}\Bigg) { S_1} ( N ) 
+ \Bigg( \,{32 \over 3 (N+2)^2}
\nonumber\\ &&
+\,{32 \over 3 (N+2)} \Bigg) { S_1} ( N+2 ) 
+ \Bigg( \,{8 \over (N-1)^2}
-\,{172 \over 3 (N-1)} \Bigg) { S_1^2} ( N-1 ) 
\nonumber\\ &&
+{1 \over N}\Bigg(-120\,{ \zeta_3 }
-104\,{ \zeta_2}
-32\,{ \beta^{(2)}} ( N+1 ) 
-36
+64\,{ A_3} ( N) 
+32\,{ A_{18}} ( N ) 
\nonumber\\ &&
-64\,{ \beta} ( N+1 ) { \zeta_2}
-80\,{ \beta^{(1)}} ( N+1 ) \Bigg)
+ \Bigg( -{\frac {772}{3}}
+64\,{ \beta^{(1)} } ( N+1 ) 
+32\,{ \zeta_2} \Bigg) {1 \over N^2}
+ \Bigg( \,{24 \over N^2}
\nonumber\\ &&
+\,{48 \over N} \Bigg) { S_1^2} ( N ) 
+ \Bigg( -\,{8 \over (N+1)}
+\,{12 \over (N+1)^2}\Bigg) { S_1^2} ( N+1 ) 
+ \Bigg( -\,{364 \over 3 (N-1)}
\nonumber\\ &&
-\,{88 \over (N-1)^2} \Bigg) { S_2} ( N-1 ) 
+ \Bigg( -\,{16 \over (N+1)}
-\,{36 \over (N+1)^2} \Bigg) { S_2} ( N+1 ) \Bigg]
\nonumber\\
%\end{eqnarray}
%%%%%%%%%%%%%%%%%%%%%%%%%%%%%%%%%%%%%%%%%%%%%%%%%%%%%%%%%%%%%%
%\begin{eqnarray}
%{ c1gCf2}
&& +C_F^2 \Biggl[
-\,{88 \over N^4}
-\,{64 \over N^3}
+\,{44 \over (N+1)^4}
+\,{166 \over (N+1)^3}
+\,{262 \over 5 (N+1)^2}
+\,{154 \over 5 (N+1)}
\nonumber\\ &&
+\,{16 \over 5 (N+2)^2}
-\,{16 \over 5 (N+2)}
-\,{16 \over 5 (N+3)^3}
+\,{316 \over 5 (N-1)}
-\,{144 \over 5 (N-1)^2}
\nonumber\\ &&
+24\,{\frac {{ \zeta_2}}{{(N+1)}^{2}}}
-12\,{\frac {{ \zeta_2}}{(N+1)}}
+{8 \over 5}\,{\frac {{ \zeta_2}}{(N+3)}}
+96\,{\frac {{ \zeta_2}}{(N-1)}}
+{\frac {32}{5}} \,{\frac {{ \zeta_2}}{(N-2)}}
\nonumber\\ &&
+{\frac {64}{5}}\,{\frac {{ \beta^{(1)}} ( N-1 ) }{(N-2)}}
-{\frac {16}{5}}\,{\frac {{ \beta^{(1)}} ( 4+N ) }{(N+3)}}
-16\,{\frac {{ \beta^{(2)}} ( N+2 ) }{(N+1)}}
-64\,{\frac {{ A_3} ( N+1 ) }{(N+1)}}
\nonumber\\ &&
+64\,{\frac {{ \beta} ( N+2 ) { \zeta_2}}{(N+1)}}
-32\,{\frac {{ \zeta_3}}{(N+1)}}
+{\frac {128}{3}}\,{\frac {{ S_3} ( N+1 ) }{(N+1)}}
+96\,{ \frac {{ \beta^{(1)}} ( N+2 ) }{(N+1)}}
+16\,{\frac {{ A_{18}} ( N+1) }{(N+1)}}
\nonumber\\ &&
-{\frac {256}{3}}\,{\frac {{ S_3} ( N ) }{N}}
-{\frac { 20}{3}}\,{\frac {{ S_1^3} ( N+1 ) }{(N+1)}}
+{\frac {40}{3}}\,{\frac {{ S_1^3} ( N ) }{N}}
+64\,{\frac {{ \zeta_2}}{{(N-1)}^{2}}}
+16\,{\frac {{ \zeta_3}}{(N-1)}}
\nonumber\\ &&
+{\frac {160}{3}}\,{\frac {{ S_3} ( N-1 ) }{(N-1)}}
-16\,{ \frac {{ \beta^{(2)}} ( N ) }{(N-1)}}
-64\,{\frac {{ A_3} ( N-1) }{(N-1)}}
+32\,{\frac {{ A_{18}} ( N-1 ) }{(N-1)}}
-{\frac {40}{3} }\,{\frac {{ S_1^3} ( N-1 ) }{(N-1)}}
\nonumber\\ &&
+64\,{\frac {{ \beta} ( N) { \zeta_2}}{(N-1)}}
+ \Bigg( -\,{48 \over (N-1)}
-\,{48 \over (N-1)^2} \Bigg) { S_1^2} ( N-1 ) 
+ \Bigg( -\,{20 \over (N+1)^2}
\nonumber\\ &&
-\,{14 \over (N+1)} \Bigg) { S_1^2} ( N+1 ) 
+ \Bigg( \,{40 \over N^2}
+\,{56 \over N} \Bigg) { S_1^2} ( N) 
+ \Bigg( -\,{8 \over N^2}
+\,{8 \over N}
%\nonumber\\ &&
\nonumber\\
%\end{eqnarray}\begin{eqnarray}
&&
-16\,{\frac {{ S_1} ( N-1 ) }{(N-1)}}
+16\,{\frac {{ S_1} ( N ) }{N}}
+\,{4 \over (N+1)^2}
-8\,{\frac {{ S_1} ( N+1 ) }{(N+1)}}
-\,{2 \over (N+1)} \Bigg) \log^2\left({Q^2 \over \mu^2}\right )
\nonumber\\ &&
+ \Bigg(  \Bigg( \,{48 \over (N-1)}
+\,{64 \over (N-1)^2} \Bigg) { S_1} ( N-1 ) 
+ \Bigg( \,{24 \over (N+1)^2}
+\,{12 \over (N+1)} \Bigg) { S_1} ( N+1 ) 
\nonumber\\ &&
+ \Bigg( -\,{64 \over N}
- \,{48 \over N^2} \Bigg) { S_1} ( N ) 
+32\,{\frac {{ S_1^2} ( N-1) }{(N-1)}}
+16\,{\frac {{ S_1^2} ( N+1 ) }{(N+1)}}
-32\,{\frac {{ S_1^2} ( N ) }{N}}
\nonumber\\ &&
+32\,{\frac {{ S_2} ( N-1 ) }{(N-1)}}
+16\, {\frac {{ S_2} ( N+1 ) }{(N+1)}}
-32\,{\frac {{ S_2} ( N ) }{N}}
-\,{24 \over (N+1)^3}
+\,{8 \over (N+1)}
\nonumber\\ &&
-\,{20 \over (N+1)^2}
+\,{8 \over (N-1)}
-64\,{\frac {{ \zeta_2}}{(N-1)}}
-32\,{\frac {{ \zeta_2}}{(N+1)}}
+{\frac {-20+64\,{ \zeta_2}}{N}}
+ \,{16 \over N^2}
\nonumber\\ &&
+\,{48 \over N^3} \Bigg) \log\left({Q^2 \over \mu^2}\right )
+ \Bigg( -{\frac {604}{5}}
-80\,{ \zeta_2}
+ 64\,{ \zeta_3}
-128\,{ A_3} ( N ) 
+128\,{ \beta} ( N+1 ) { \zeta_2}
\nonumber\\ &&
-32\,{ \beta^{(2)}} ( N+1 ) 
+64\,{ \beta^{(1)}} ( N+1 ) 
-32\,{ A_{18}} ( N )  \Bigg) {1 \over N}
+ \Bigg( -128\,{ \beta^{(1)}} ( N+1) 
-{\frac {418}{5}}
\nonumber\\ &&
-112\,{ \zeta_2} \Bigg) {1 \over N^2}
+ \Bigg( -\,{48 \over (N-1)} 
-\,{112 \over (N-1)^2} \Bigg) { S_2} ( N-1 ) 
+ \Bigg( -\,{8 \over (N-1)}
\nonumber\\ &&
+64\,{ \frac {{ \beta^{(1)}} ( N ) }{(N-1)}}
-\,{64 \over (N-1)^3}
-56\,{\frac {{ S_2} ( N-1 ) }{(N-1)}}
-\,{96 \over (N-1)^2} \Bigg) { S_1} ( N-1 ) 
\nonumber\\ &&
+ \Bigg( -\,{52 \over (N+1)^2}
+\,{2 \over (N+1)} \Bigg) { S_2} ( N+1 ) 
+ \Bigg( -\,{12 \over (N+1)^2}
+\,{24 \over (N+1)}
\nonumber\\ &&
+64\,{\frac {{ \beta^{(1)}} ( N+2 ) } {(N+1)}}
-28\,{\frac {{ S_2} ( N+1 ) }{(N+1)}}
-16\,{\frac {{ \zeta_2}}{(N+1)}}
-\,{8 \over (N+1)^3} \Bigg) { S_1} ( N+1 ) 
\nonumber\\ &&
+ \Bigg( 56\,{\frac {{ S_2 } ( N ) }{N}}
+{\frac {32\,{ \zeta_2}
+128\,{ \beta^{(1)}} ( N+1 ) }{N}}
+\,{80 \over N^2}
+\,{16 \over N^3} \Bigg) { S_1} ( N ) 
\nonumber\\ &&
+ \Bigg( \,{56 \over N}
+ \,{104 \over N^2} \Bigg) { S_2} ( N ) \Bigg]
\\
%\end{eqnarray}
%%%%%%%%%%%%%%%%%%%%%%%%%%%%%%%%%%%%%%%%%%%%%%%%%%%%%%%%%%%%%%%%%%%
%\begin{eqnarray}
%{ cLqCf2ni}
C_{L,q}^{(2),nid}&=&C_F^2 \Bigg[
-\,{20 \over N^3}
+\,{16 \over (N+1)^3}
-\,{32 \over 5(N+1)^2}
+\,{86 \over 5 (N+1)}
+\,{8 \over (N+2)^3}
\nonumber\\ &&
-\,{16 \over 5(N+2)^2}
+\,{16 \over 5(N+2)}
+\,{16 \over 5(N+3)^3}
+\,{24 \over 5(N-1)}
+\,{24 \over 5(N-1)^2}
\nonumber\\ &&
-8\,{\frac {{ \zeta_2}}{(N+1)}}
-4\,{\frac {{\zeta_2}}{(N+2)}}
-{8 \over 5}\,{\frac {{\zeta_2}}{(N+3)}}
-8\,{\frac {{\zeta_2}}{(N-1)}}
-{\frac {12}{5}}\,{\frac {{ \zeta_2}}{ (N-2)}}
-4\,{\frac {{ S_1} ( N+1 ) }{(N+1)}}
\nonumber\\ &&
-{\frac {24}{5}}\,{\frac {{ \beta^{(1)}} ( N-1 ) }{(N-2)}}
+8\,{\frac {{ \beta^{(1)}} ( N+3 ) }{(N+2)}}
+{\frac {16}{5}}\,{\frac {{ \beta^{(1)}} ( 4+N ) }{(N+3)}}
-16\,{\frac {{ \beta^{(1)}} ( N ) }{(N-1)}}
+20\,{\frac {{ S_2} ( N ) }{N}}
\nonumber\\ &&
+4\,{ \frac {{ S_1^2} ( N ) }{N}}
+16\,{\frac {{ \beta^{(1)}} ( N+2 ) }{(N+1)}}
-16\,{\frac {{ S_3} ( N ) }{N}}
+ \Bigg( \,{4 \over (N+1)}
-8\,{\frac {{ S_1} ( N ) }{N}}
+\,{2 \over N}
\nonumber\\ &&
+\,{4 \over N^2} \Bigg) \log\left({Q^2 \over \mu^2}\right )
+ \Bigg( { \frac {16\,{ \zeta_2}-14}{N}}
+\,{4 \over N^2} \Bigg) { S_1} ( N ) 
+ \Bigg( -{\frac {211}{5}}
+40\,{ \zeta_3}
+8\,{ \zeta_2} \Bigg) {1 \over N}
\nonumber\\ &&
+ \Bigg( {\frac {78}{ 5}}
-8\,{ \zeta_2}
-16\,{ \beta^{(1)}} ( N+1 )  \Bigg) {1 \over N^2}\Bigg]
\nonumber\\ &&
%\end{eqnarray}
%%%%%%%%%%%%%%%%%%%%%%%%%%%%%%%%%%%%%%%%%%%%%%%%%%%%%%%%%%%%%
%\begin{eqnarray}
%{ cLqCaCfni}
+ C_A C_F \Bigg[
\,{4 \over N^3}
-\,{8 \over (N+1)^3}
-\,{4 \over 5 (N+1)^2}
-\,{254 \over 15 (N+1)}
-\,{4 \over (N+2)^3}
+\,{8 \over 5 (N+2)^2}
\nonumber\\ &&
-\,{8 \over 5 (N+2)}
-\,{8 \over 5 (N+3)^3}
-\,{12 \over 5 (N-1)}
-\,{12 \over 5 (N-1)^2}
-{\frac {22}{3 N}} \log\left({Q^2 \over \mu^2}\right )
\nonumber\\ &&
+4\,{\frac {{ \zeta_2}}{(N+1)}}
+2\,{\frac {{ \zeta_2}}{(N+2)}}
+{4 \over 5}\,{\frac {{ \zeta_2}}{(N+3)}}
+4\,{\frac {{ \zeta_2}}{(N-1)}}
+{6 \over 5}\,{\frac {{ \zeta_2}}{(N-2)}}
+{\frac {12}{5}}\,{ \frac {{ \beta^{(1)}} ( N-1 ) }{(N-2)}}
\nonumber\\ &&
-4\,{\frac {{ \beta^{(1)}} ( N+3) }{(N+2)}}
-{8 \over 5}\,{\frac {{ \beta^{(1)}} ( 4+N ) }{(N+3)}}
+8\,{\frac {{ \beta^{(1)}} ( N ) }{(N-1)}}
-8\,{\frac {{ \beta^{(1)}} ( N+2 ) }{(N+1)} }
+8\,{\frac {{ S_3} ( N ) }{N}}
\nonumber\\ &&
+ \Bigg( {\frac {46}{3}}
-8\,{ \zeta_2}\Bigg) { S_1} ( N ) {1 \over N}
+ \Bigg( -{\frac {22}{15}}
+4\,{ \zeta_2}
+8 \,{ \beta^{(1)}} ( N+1 )  \Bigg) {1 \over N^2}
+ \Bigg( {\frac {2017}{45}}
\nonumber\\ &&
-20\,{ \zeta_3}
-4\,{ \zeta_2} \Bigg) {1 \over N} \Bigg]
\nonumber\\ &&
%\end{eqnarray}
%%%%%%%%%%%%%%%%%%%%%%%%%%%%%%%%%%%%%%%%%%%%%%%%%%%%%%%%%%%%%%%
%\begin{eqnarray}
%{ cLqnfCfTfni}
+ N_F C_F T_F  \Bigg[
-\,{8 \over 3 N^2}
+{ 8 \over 3 N}\,\log\left({Q^2 \over \mu^2}\right )
-\,{100 \over 9 N}
+\,{8 \over 3 (N+1)}
-{8 \over 3}\,{\frac {{ S_1} ( N ) }{N}} \Bigg]
\\
%\end{eqnarray}
%%%%%%%%%%%%%%%%%%%%%%%%%%%%%%%%%%%%%%%%%%%%%%%%%%%%%%%%%%%%%%%
%\begin{eqnarray}
%{ cLqCf2id}
C_{L,q}^{(2),id}&=&C_F^2 \Bigg[
32\,{\frac {{ S_1} ( N ) { \beta^{(1)}} ( N+1 ) }{N}}
-\,{104 \over 5 (N+1)}
+8\,{\frac {{ \zeta_2}}{(N-1)}}
-{\frac {12}{5}}\,{\frac {{ \zeta_2 }}{(N-2)}}
\nonumber\\ &&
+16\,{\frac {{ \beta^{(1)}} ( N+2 ) }{(N+1)}}
-8\,{\frac {{ \beta^{(1)}} ( N+3 ) }{(N+2)}}
+{\frac {16}{5}}\,{\frac {{ \beta^{(1)}} ( 4+N ) } {(N+3)}}
-{\frac {24}{5}}\,{\frac {{ \beta^{(1)}} ( N-1 ) }{(N-2)}}
+4\,{\frac {{ \zeta_2}}{(N+2)}}
\nonumber\\ &&
-8\,{\frac {{ \zeta_2}}{(N+1)}}
-{8 \over 5}\,{\frac {{ \zeta_2}}{(N+3)}}
- \,{8 \over (N+2)^3}
-\,{16 \over 5 (N+2)^2}
+\,{16 \over 5 (N+2)}
\nonumber\\ &&
+\,{16 \over 5 (N+3)^3}
+\,{24 \over 5 (N-1)}
+\,{24 \over 5 (N-1)^2}
+\,{16 \over (N+1)^3}
+\,{8 \over 5 (N+1)^2}
\nonumber\\ &&
+16\,{\frac {{ \beta^{(1)}} ( N ) }{(N-1)}}
+ \Bigg( 32\,{ \beta} ( N+1 ) { \zeta_2}
+16\,{ \beta^{(1)}} ( N+1) 
+{\frac {64}{5}}
-32\,{ A_3} ( N ) 
\nonumber\\ &&
-8\,{ \beta^{(2)}} ( N+1)  \Bigg) {1 \over N}
+ \Bigg( -8\,{ \zeta_2}
-{\frac {112}{5}}
-16\,{ \beta^{(1)}} ( N+1 )  \Bigg) {1 \over N^2}
+\,{8 \over N^3} \Bigg]
\nonumber\\ &&
%\end{eqnarray}
%%%%%%%%%%%%%%%%%%%%%%%%%%%%%%%%%%%%%%%%%%%%%%%%%%%%%%%%%%%%%
%\begin{eqnarray}
%{ cLqCaCfid}
+ C_A C_F \Bigg[
-16\,{\frac {{ S_1} ( N ) { \beta^{(1)}} ( N+1 ) }{N}}
+{\frac {12} {5}}\,{\frac {{ \beta^{(1)}} ( N-1 ) }{(N-2)}}
+{6 \over 5}\,{\frac {{ \zeta_2}}{(N-2)} }
-4\,{\frac {{ \zeta_2}}{(N-1)}}
\nonumber\\ &&
-8\,{\frac {{ \beta^{(1)}} ( N ) }{(N-1)}}
+4\, {\frac {{ \beta^{(1)}} ( N+3 ) }{(N+2)}}
+4\,{\frac {{ \zeta_2}}{(N+1)}}
+{4 \over 5}\,{ \frac {{ \zeta_2}}{(N+3)}}
-{\frac {12}{5}}\,{1 \over (N-1)^2}
\nonumber\\ &&
-\,{8 \over 5 (N+3)^3}
-\,{12 \over 5 (N-1)}
-\,{4 \over 5 (N+1)^2}
+\,{8 \over 5 (N+2)^2}
-\,{8 \over 5 (N+2)}
\nonumber\\ &&
+\,{4 \over (N+2)^3}
+\,{52 \over 5 (N+1)}
-\,{8 \over (N+1)^3}
-8\,{\frac {{ \beta^{(1)}} ( N+2) }{(N+1)}}
-{8 \over 5}\,{\frac {{ \beta^{(1)}} ( 4+N ) }{(N+3)}}
\nonumber\\ &&
-2\,{\frac {{ \zeta_2}}{(N+2)}}
+ \Bigg( -{\frac {32}{5}}
+16\,{ A_3} ( N ) 
-8\,{ \beta^{(1)}} ( N+1 ) 
-16\,{ \beta} ( N+1 ) { \zeta_2}
\nonumber\\ &&
+4\,{ \beta^{(2)}} ( N+1 )  \Bigg) {1 \over N}
+ \Bigg( {\frac {56}{5}}
+8\,{ \beta^{(1)}} ( N+1) 
+4\,{ \zeta_2} \Bigg) {1 \over N^2}
-\,{4 \over N^3} \Bigg]
\\
%\end{eqnarray}
%%%%%%%%%%%%%%%%%%%%%%%%%%%%%%%%%%%%%%%%%%%%%%%%%%%%%%%%%%%%%
%\begin{eqnarray}
%{ cLqnfCfTfPS}
C_{L,q}^{{\rm PS}(2)}&=& N_F C_F T_F  \Bigg[
 \Bigg( -\,{16 \over (N+1)}
-\,{16 \over N^2}
+\,{32 \over 3 (N-1)}
+\,{16 \over 3 (N+2)} \Bigg) \log\left({Q^2 \over \mu^2}\right )
\nonumber\\ &&
+\,{48 \over N^3}
+16\,{\frac {{ S_1} ( N+1 ) }{(N+1)}}
+ \,{32 \over N^2}
-\,{112 \over 3 N}
+\,{32 \over (N+1)^2}
-\,{16 \over (N+2)}
\nonumber\\ &&
+\,{208 \over 3 (N+1)}
-\,{16 \over 3 (N+2)^2}
-\,{16 \over (N-1)}
-{\frac {32}{3}}\,{\frac {{ S_1} ( N-1 ) }{(N-1)}}
-\,{64 \over 3 (N-1)^2}
\nonumber\\ &&
-{16 \over 3}\,{\frac {{ S_1} ( N+2 ) }{(N+2)}}
+16\,{\frac {{ S_1} ( N ) }{{N}^{2}}}\Bigg]
%\end{eqnarray}
\\
%%%%%%%%%%%%%%%%%%%%%%%%%%%%%%%%%%%%%%%%%%%%%%%%%%%%%%%%%%%%%%%
%\begin{eqnarray}
%{ cLgCaCf}
C_{L,g}^{{\rm S}(2)}&=&C_A C_F \Bigg[
 \Bigg( \,{32 \over N^2}
+32\,{\frac {{ S_1} ( N ) }{N}}
-{\frac {272}{3}}\,{1 \over (N-1)}
+\,{32 \over (N-1)^2}
-\,{16 \over 3 (N+2)}
\nonumber\\ &&
+\,{16 \over (N+1)}
+\,{80 \over N}
-32\,{ \frac {{ S_1} ( N-1 ) }{(N-1)}} \Bigg) \log\left({Q^2 \over \mu^2}\right )
+{\frac {464}{3}}\,{ \frac {{ S_1} ( N-1 ) }{(N-1)}}
-16\,{\frac {{ S_1} ( N+1 ) } {(N+1)}}
\nonumber\\ &&
+ \Bigg( -\,{64 \over N^2}
-\,{144 \over N} \Bigg) { S_1} ( N ) 
+{16 \over 3}\, {\frac {{ S_1} ( N+2 ) }{(N+2)}}
+16\,{\frac {{ S_1^2} ( N-1) }{(N-1)}}
-16\,{\frac {{ S_1^2} ( N ) }{N}}
\nonumber\\ &&
+80\,{\frac {{ S_2} ( N-1 ) }{(N-1)}}
-80\,{\frac {{ S_2} ( N ) }{N}}
+\,{352 \over 3 (N-1)^2}
+\,{16 \over 3 (N+2)^2}
-\,{128 \over (N-1)^3}
\nonumber\\ &&
+\,{448 \over 3 (N-1)} 
-32\,{\frac {{ \beta^{(1)}} ( N ) }{(N-1)}}
+\,{32 \over 3 (N+2)}
+16\,{ \frac {{ \zeta_2}}{(N-1)}}
-\,{160 \over 3 (N+1)}
\nonumber\\ &&
-\,{32 \over (N+1)^2}
+ \Bigg( - 32\,{ \beta^{(1)}} ( N+1 ) 
-{\frac {320}{3}}
-16\,{ \zeta_2} \Bigg) {1 \over N}
- \,{112 \over N^2}
-\,{96 \over N^3} \Bigg]
\nonumber\\ &&
%\end{eqnarray}
%%%%%%%%%%%%%%%%%%%%%%%%%%%%%%%%%%%%%%%%%%%%%%%%%%%%%%%%%%%%%%%%%%%
%\begin{eqnarray}
%{ cLgCf2}
+ C_F^2 \Bigg[
 \Bigg( -\,{8 \over (N+1)}
-\,{8 \over N}
-\,{16 \over N^2}
+\,{16 \over (N-1)} \Bigg)  \log\left({Q^2 \over \mu^2}\right )
- 32\,{\frac {{ S_1} ( N-1 ) }{(N-1)}}
\nonumber\\ &&
+8\,{\frac {{ S_1} ( N+1) }{(N+1)}}
+ \Bigg( \,{24 \over N}
+\,{16 \over N^2} \Bigg) { S_1} ( N) 
-{\frac {64}{5}}\,{\frac {{ \beta^{(1)}} ( N-1 ) }{(N-2)}}
-\,{96 \over 5 (N-1)^2}
\nonumber\\ &&
+\,{224 \over 15(N+1)^2}
+\,{248 \over 15 (N+1)}
+\,{32 \over 15 (N+2)^2}
-\,{32 \over 15 (N+2)}
-\,{32 \over 15 (N+3)^3}
\nonumber\\ &&
-{\frac {32}{5}}\,{\frac {{ \zeta_2}}{(N-2)}}
-{\frac {32}{15}}\,{\frac {{ \beta^{(1)}} ( 4+N ) }{(N+3)}}
-\,{96 \over 5 (N-1)}
+{\frac {16}{15}}\,{ \frac {{ \zeta_2}}{(N+3)}}
+ \Bigg( {\frac {24}{5}}
+{16 \over 3}\,{ \zeta_2}
\nonumber\\ &&
+{\frac {32}{3}} \,{ \beta^{(1)}} ( N+1 )  \Bigg) {1 \over N}
+\,{8 \over 5 N^2}
+\,{48 \over N^3}\Bigg]
\end{eqnarray}
%%%%%%%%%%%%%%%%%%%%%%%%%%%%%%%%%%%%%%%%%%%%%%%%%%%%%%%%%%%%%
\vspace*{2mm}
\begin{eqnarray}
\lefteqn{\C_{A,q}^{{\rm NS},nid,(2)} -   
\C_{T,q}^{{\rm NS},nid,(2)} =}\nonumber \\   
&&
        C_A  C_F \Bigg( 
         \log\left({Q^2\over\mu^2}\right)    \Bigg(
           {22 \over 3  N}
         - {22 \over 3  (N+1)}
         \Bigg )
       +     \Bigg(
          - {8  \zeta_2  S_1(N) \over N}
          + {8  \zeta_2  S_1(N+1) \over (N+1)}
\nonumber\\&&
          + {4  \zeta_2 \over N^2}
          - {4  \zeta_2 \over (N+1)^2}
          - {2  \zeta_2 \over (N+2)}
          - {6  \zeta_2 \over 5 (N+3)}
          + {2  \zeta_2 \over (N-1)}
          + {6  \zeta_2 \over 5 (N-2)}
          - {20  \zeta_3 \over N}
\nonumber\\&&
          + {20  \zeta_3 \over (N+1)}
          + {50  S_1(N) \over 3 N}
          + {8  S_3(N)\over  N}
          - {50  S_1(N+1)\over  3 (N+1)}
          - {8  S_3(N+1) \over (N+1)}
\nonumber\\&&
          + {12  \beta^{(1)}(N-1) \over 5 (N-2)}
          + {4  \beta^{(1)}(N) \over (N-1)}
          + {8  \beta^{(1)}(N+1) \over N^2}
          - {8  \beta^{(1)}(N+1) \over N}
\nonumber\\&&
          - {8  \beta^{(1)}(N+2) \over (N+1)^2}
          - {8  \beta^{(1)}(N+2) \over (N+1)}
          + {4  \beta^{(1)}(N+3) \over (N+2)}
          + {12  \beta^{(1)}(N+4) \over 5 (N+3)}
\nonumber\\&&
          - {4 \over N^3}
          - {182 \over 15 N^2}
          - {823 \over 45 N}
          - {4 \over (N+1)^3}
          + {398 \over 15 (N+1)^2}
          + {823 \over 45 (N+1)}
          + {4 \over (N+2)^3}
\nonumber\\&&
          - {12 \over 5 (N+2)^2}
          + {12 \over 5 (N+2)}
          + {12 \over 5 (N+3)^3}
          - {12 \over 5 (N-1)^2}
          - {12 \over 5 (N-1)}
          \Bigg)
\Bigg)
\nonumber\\&&
       + N_F  C_F  T_F 
\Bigg( 
        \log\left({Q^2\over\mu^2}\right)    \Bigg(\Bigg(
          - {8 \over 3 N}
          + {8 \over 3 (N+1)}
          \Bigg)
       +     \Bigg(
           {8  S_1(N)\over  3 N}
          - {8  S_1(N+1) \over 3 (N+1)}
\nonumber\\&&
%          - {8 \over 3 N^2}
          + {8 \over 3 N^2}
          + {76 \over 9 N}
%          + {8 \over 3 (N+1)^2}
          - {8 \over 3 (N+1)^2}
          - {76 \over 9 (N+1)}\Bigg)
\Bigg)
       + C_F^2  \Bigg(
         \log\left({Q^2\over\mu^2}\right)   \Bigg (
           {8  S_1(N) \over N}
\nonumber\\&&
          - {8  S_1(N+1) \over (N+1)}
          - {4 \over N^2}
          + {2 \over N}
          + {4 \over (N+1)^2}
          - {2 \over (N+1)}
          \Bigg)
       +    \Bigg(
           {16  \zeta_2  S_1(N)\over  N}
\nonumber\\&&
          - {16  \zeta_2  S_1(N+1) \over (N+1)}
          - {8  \zeta_2 \over N^2}
          + {8  \zeta_2 \over (N+1)^2}
          + {4  \zeta_2 \over (N+2)}
          + {12  \zeta_2 \over 5 (N+3)}
          - {4  \zeta_2 \over (N-1)}
\nonumber\\&&
          - {12  \zeta_2 \over 5 (N-2)}
          + {40  \zeta_3 \over N}
          - {40  \zeta_3 \over (N+1)}
          - {4  S_1(N)\over  N^2}
          - {50  S_1(N)\over  N}
          - {4  S_1^2(N)\over  N}
\nonumber\\&&
          - {20  S_2(N)\over  N}
          - {16  S_3(N)\over  N}
          + {4  S_1(N+1) \over (N+1)^2}
          + {50  S_1(N+1) \over (N+1)}
          + {4  S_1^2(N+1) \over (N+1)}
\nonumber\\&&
          + {20  S_2(N+1) \over (N+1)}
          + {16  S_3(N+1) \over (N+1)}
          - {24  \beta^{(1)}(N-1)\over  5 (N-2)}
          - {8  \beta^{(1)}(N) \over (N-1)}
\nonumber\\&&
          - {16  \beta^{(1)}(N+1)\over  N^2}
          + {16  \beta^{(1)}(N+1)\over  N}
          + {16  \beta^{(1)}(N+2) \over (N+1)^2}
          + {16  \beta^{(1)}(N+2) \over (N+1)}
%\nonumber\\&&
\nonumber\\
%\end{eqnarray}\begin{eqnarray}
&&
          - {8  \beta^{(1)}(N+3) \over (N+2)}
          - {24  \beta^{(1)}(N+4) \over 5 (N+3)}
          + {20 \over N^3}
          - {2 \over 5 N^2}
          + {9 \over 5 N}
          - {4\over  (N+1)^3}
\nonumber\\&&
          - {202 \over 5 (N+1)^2}
          - {9\over  5 (N+1)}
          - {8 \over (N+2)^3}
          + {24\over  5 (N+2)^2}
          - {24 \over 5 (N+2)}
\nonumber\\&&
          - {24\over  5 (N+3)^3}
          + {24\over 5 (N-1)}
          + {24 \over 5 (N-1)^2}
          \Bigg)
\Bigg)
\\
\lefteqn{   
\C_{A,q}^{{\rm NS},id,(2)} -   
\C_{T,q}^{{\rm NS},id,(2)} =}\nonumber\\    
&& \Bigg(C_F^2 -{1 \over 2} C_A C_F\Bigg)
       \Bigg( {32 \zeta_2 \beta(N+1) \over N}
          + {32 \zeta_2 \beta(N+2) \over (N+1)}
          - {8 \zeta_2 \over N^2}
          + {8 \zeta_2 \over N}
          - {8 \zeta_2 \over (N+1)^2}
\nonumber\\&&
          - {8 \zeta_2 \over (N+1)}
          - {4 \zeta_2 \over (N+2)}
          + {12 \zeta_2 \over 5 (N+3)}
          + {4 \zeta_2 \over (N-1)}
          - {12 \zeta_2 \over 5 (N-2)}
          + {32 S_1(N) \beta^{(1)}(N+1) \over N}
\nonumber\\&&
          + {32 S_1(N+1) \beta^{(1)}(N+2) \over (N+1)}
          - {32 A_3(N) \over N}
          - {32 A_3(N+1) \over (N+1)}
          - {24 \beta^{(1)}(N-1) \over 5 (N-2)}
\nonumber\\&&
          + {8 \beta^{(1)}(N) \over (N-1)}
          - {16 \beta^{(1)}(N+1) \over N^2}
          - {8 \beta^{(2)}(N+1) \over N}
          - {16 \beta^{(1)}(N+2) \over (N+1)^2}
\nonumber\\&&
          - {8 \beta^{(2)}(N+2) \over (N+1)}
          + {8 \beta^{(1)}(N+3) \over (N+2)}
          - {24 \beta^{(1)}(N+4) \over 5 (N+3)}
          - {8 \over N^3}
          - {72 \over 5 N^2}
\nonumber\\&&
          + {104 \over 5 N}
          + {8 \over (N+1)^3}
          - {72 \over 5 (N+1)^2}
          - {104 \over 5 (N+1)}
          + {8 \over (N+2)^3}
          + {24 \over 5 (N+2)^2}
\nonumber\\&&
          - {24 \over 5 (N+2)}
          - {24 \over 5 (N+3)^3}
          + {24 \over 5 (N-1)^2}
          + {24 \over 5 (N-1)}\Bigg)
\end{eqnarray}

\vspace{2mm}\noindent
The functions $\C_{T,q}^{{\rm NS}, nid(id), (2)}(N)$ were given in (\ref{eqA5}) summing over 
the color factors.

%\end{appendix}
%%%%%%%%%%%%%%%%%%%%%%%%%%%%%%%%%%%%%%%%%%%%%%%%%%%%%%%%%%%%%
\newpage
%\begin{appendix}
%\setcounter{equation}{0}
\section{Polarized Time--like Coefficient Functions}

\label{sec-B}
\renewcommand{\theequation}{\thesection.\arabic{equation}}
\setcounter{equation}{0}
%\section{B}

\vspace{1mm}\noindent
In this appendix, we present the Mellin moments of the polarized time--like  coefficient functions to
$O(\alpha_s^2)$.

%%%%%%%%%%%%%%%%%%%%%%%%%%%%%%%%%%%%%%%%%%%%%%%%%%%%%%%%%%%%%%%%%%%%%%%%%%%%%%%%%%%%%%%%%%%%%%%%%%%
%\setcounter{equation}{46}  
\begin{eqnarray}
\Delta \C_{T,a}=\Delta C_{T,a}^{(0)}+a_s~ \Delta C_{T,a}^{(1)}+a_s^2~\Delta  C_{T,a}^{(2)}
\end{eqnarray}

\vspace{1mm}\noindent
The individual contributions are
\begin{eqnarray}
%--------------------------------------------------------------------------------------------------
%{ g1qCfnfTf}
\Delta C_{1,q}^{{\rm PS}(2)}&=& N_F C_F  T_F \Bigg[
 \Bigg( \,{20 \over N}
-\,{20 \over (N+1)}
-\,{8 \over N^2}
-\,{8 \over (N+1)^2} \Bigg) \log^2\left({Q^2 \over \mu^2}\right )
+ \Bigg(  (\,{40 \over (N+1)}
\nonumber\\ &&
+\,{16 \over (N+1)^2} ) { S_1} ( N+1) 
+ \Bigg( -\,{40 \over N}
+\,{16 \over N^2} \Bigg) { S_1} ( N ) 
+{136\over (N+1)}
+\,{88 \over (N+1)^2}
\nonumber\\ &&
-\,{136 \over N}
+\,{48 \over N^3}
+\,{48 \over (N+1)^3}
-\,{8 \over N^2}\Bigg) \log\left({Q^2 \over \mu^2}\right )
+ \Bigg( -\,{88 \over (N+1)^2}
\nonumber\\ &&
-\,{136 \over (N+1)}
-\,{48 \over (N+1)^3}\Bigg) { S_1} ( N+1 ) 
+ \Bigg( \,{136 \over N}
-\,{48 \over N^3}
+\,{8 \over N^2}\Bigg) { S_1} ( N ) 
\nonumber\\ &&
+ \Bigg( -\,{20 \over (N+1)}
-\,{8 \over (N+1)^2} \Bigg) { S_1^2} ( N+1 ) 
+ \Bigg( \,{20 \over N}
-\,{8 \over N^2} \Bigg) { S_1^2} ( N ) 
+ \Bigg( -\,{20 \over (N+1)}
\nonumber\\ &&
-\,{8 \over (N+1)^2} \Bigg) { S_2} ( N +1 ) 
+ \Bigg( \,{20 \over N}
-\,{8 \over N^2} \Bigg) { S_2} ( N ) 
-\,{284 \over 3 (N+1)^2}
+56\,{\frac {{ \zeta_2}}{(N+1)}}
\nonumber\\ &&
-{16 \over 3}\,{\frac {{ \zeta_2}}{(N+2)}}
-\,{424 \over 3 (N+1)}
+{\frac {32}{3}}\,{\frac {{ \beta^{(1)}} ( N+3 ) }{(N+2)}}
+{\frac {32}{3}}\,{\frac {{ \beta^{(1)}} ( N ) }{(N-1)}}
-\,{100 \over (N+1)^3}
\nonumber\\ &&
+{16 \over 3}\,{\frac {{ \zeta_2}}{(N-1)}}
+\,{32 \over 3 (N+2)^3}
-\,{88 \over (N+1)^4}
+16\,{\frac {{ \zeta_2}}{{(N+1)}^{2}}}
+32\,{\frac {{ \beta^{(1)}} ( N+2 ) }{(N+1)}}
\nonumber\\ &&
+ \Bigg( {\frac {424}{3}}
-56\,{ \zeta_2}
+32\,{ \beta^{(1)}} ( N+1 )  \Bigg) {1 \over N}
+ \Bigg( {\frac {340}{3}}
+16\,{ \zeta_2}\Bigg) {1 \over N^2}
-\,{44 \over N^3}
-\,{88 \over N^4}
\Bigg]
\end{eqnarray}
%%%%%%%%%%%%%%%%%%%%%%%%%%%%%%%%%%%%%%%%%%%%%%%%%%%%%%%%%%%%%%%%%
\begin{eqnarray}
%{ g1gCaCf}
\Delta C_{1,g}^{(2)}&=& C_A C_F \Bigg[
\,{496 \over N^4}
+\,{144 \over N^3}
+\,{88 \over (N+1)^4}
+\,{76 \over (N+1)^3}
+\,{128 \over 3 (N+1)^2}
\nonumber\\ &&
+\,{76 \over 3 (N+1)}
-\,{32 \over 3 (N+2)^3}
+16\,{\frac {{ \zeta_2}}{{(N+1)}^{2}}}
-112\,{\frac {{ \zeta_2}}{(N+1)}}
+{16 \over 3}\,{\frac {{ \zeta_2}}{(N+2)}}
\nonumber\\ &&
-{\frac {88}{3}}\,{\frac {{ \zeta_2}}{(N-1)}}
-{\frac {32}{3}}\,{\frac {{ \beta^{(1)}} ( N+3 ) }{(N+2)}}
-{\frac {176}{3}}\,{\frac {{ \beta^{(1)}} ( N ) }{ (N-1)}}
-16\,{\frac {{ \beta^{(2)}} ( N+2 ) }{(N+1)}}
+32\,{\frac {{ A_3} ( N+1 ) }{(N+1)}}
\nonumber\\ &&
-32\,{\frac {{ \beta} ( N+2 ) { \zeta_2}}{(N+1)}}
-60\,{\frac {{ \zeta_3}}{(N+1)}}
+{\frac {200}{3}}\,{\frac {{ S_3} ( N+1) }{(N+1)}}
+32\,{\frac {{ \beta^{(1)}} ( N+2 ) }{{(N+1)}^{2}}}
\nonumber\\ &&
-32\,{ \frac {{ \beta^{(1)}} ( N+2 ) }{(N+1)}}
+16\,{\frac {{ A_{18}} ( N+1) }{(N+1)}}
+ \Bigg( -16\,{\frac {{ \beta^{(1)}} ( N+2 ) }{(N+1)}}
+\,{64 \over (N+1)^3}
\nonumber\\ &&
+\,{136 \over (N+1)^2}
+36\,{\frac {{ S_2} ( N+1 ) }{(N+1)}}
-40\,{ \frac {{ \zeta_2}}{(N+1)}}
+\,{240 \over (N+1)} \Bigg) { S_1} ( N+1 ) 
\nonumber\\ &&
+ \Bigg( \,{12 \over (N+1)^2}
+\,{48 \over (N+1)} \Bigg) { S_1^2} ( N+1 ) 
-{ \frac {400}{3}}\,{\frac {{ S_3} ( N ) }{N}}
+{4 \over 3}\,{\frac {{ S_1^3} ( N+1 ) }{(N+1)}}
-{8 \over 3}\,{\frac {{ S_1^3} ( N ) }{N}}
\nonumber\\ &&
+16\,{ \frac {{ \zeta_2}}{{(N-1)}^{2}}}
-40\,{\frac {{ \zeta_3}}{(N-1)}}
+16\,{\frac {{ S_3 } ( N-1 ) }{(N-1)}}
+32\,{\frac {{ \beta^{(1)}} ( N ) }{{(N-1)}^{2}}} 
+8\,{\frac {{ \beta^{(2)}} ( N ) }{(N-1)}}
\nonumber\\ &&
+32\,{\frac {{ A_3} ( N-1) }{(N-1)}}
-32\,{\frac {{ \beta} ( N ) { \zeta_2}}{(N-1)}}
+ \Bigg( \,{24 \over N^2}
-\,{56 \over N} \Bigg) { S_1^2} ( N ) 
+ \Bigg( \,{8 \over (N+1)^2}
+\,{32 \over N^2}
\nonumber\\ &&
+\,{48 \over (N+1)}
-16\,{\frac {{ S_1} ( N ) }{N}}
-\, {48 \over N}
+8\,{\frac {{ S_1} ( N+1 ) }{(N+1)}} \Bigg) \log^2\left({Q^2 \over \mu^2}\right )
+ \Bigg( -{\frac {124}{3}}
\nonumber\\ &&
-32\,{ \beta^{(2)}} ( N+1 ) 
+128\,{ \zeta_2}
+120\,{ \zeta_3}
-64\,{ \beta^{(1)}} ( N+1 ) 
-64\,{ \beta} ( N+1 ) { \zeta_2}
+64\,{ A_3} ( N ) 
\nonumber\\ &&
-32\,{ A_{18}} ( N )  \Bigg) {1 \over N}
+ \Bigg( -{\frac {844}{3}}
-128\,{ \zeta_2}+64\,{ \beta^{(1)}} ( N+1 ) \Bigg) {1 \over N^2}
+ \Bigg( -72\,{\frac {{ S_2} ( N ) }{N}}
\nonumber\\ &&
+{1 \over N}\Bigg(80\,{ \zeta_2}
-276
-32\,{ \beta^{(1)}} ( N+1 ) )
-\,{48 \over N^2}
+\,{192 \over N^3}\Bigg) { S_1} ( N ) 
+ \Bigg(  \Bigg( -\,{112 \over (N+1)}
\nonumber\\ &&
-\,{32 \over (N+1)^2} \Bigg) { S_1} ( N+1 ) 
+ \Bigg( -\,{32 \over N^2}
+\,{128 \over N} \Bigg) { S_1} ( N ) 
-8\,{\frac {{ S_1^2} ( N+1 ) }{(N+1)}}
+16\,{ \frac {{ S_1^2} ( N ) }{N}}
\nonumber\\ &&
-40\,{\frac {{ S_2} ( N+1 ) }{(N+1)}}
+80\,{\frac {{ S_2} ( N ) }{N}}
-16\,{\frac {{ \beta^{(1)}} ( N+2) }{(N+1)}}
-\,{224 \over (N+1)}
-8\,{\frac {{ \zeta_2}}{(N+1)}}
\nonumber
\end{eqnarray}\begin{eqnarray}
%\nonumber\\ 
&&
-\,{136 \over (N+1)^2} 
-\,{48 \over (N+1)^3}
+{\frac {232-32\,{ \beta^{(1)}} ( N+1 ) 
+16\,{ \zeta_2}}{N}}
+ \,{32 \over N^2}
\nonumber\\ &&
-\,{224 \over N^3} \Bigg) \log\left({Q^2 \over \mu^2}\right )
+ \Bigg( \,{128 \over (N+1)}
+\,{60 \over (N+1)^2} \Bigg) { S_2} ( N+1 ) 
+ \Bigg( \,{8 \over (N-1)}
\nonumber\\ &&
-16\,{\frac {{ \zeta_2} }{(N-1)}}
-32\,{\frac {{ \beta^{(1)}} ( N ) }{(N-1)}} \Bigg) { S_1} ( N- 1 ) 
+ \Bigg( -\,{72 \over N^2}
-\,{168 \over N} \Bigg) { S_2} ( N ) 
\Bigg]
\nonumber\\
%\end{eqnarray}
%%%%%%%%%%%%%%%%%%%%%%%%%%%%%%%%%%%%%%%%%%%%%%%%%%%%%%%%%%%%%%%%
%\begin{eqnarray}
%{ g1gCf2}
&&
+ C_F^2 \Bigg[
\,{88 \over N^4}
+\,{256 \over N^3}
-\,{44 \over (N+1)^4}
-\,{86 \over (N+1)^3}
-\,{196 \over 3 (N+1)^2}
-\,{110 \over 3 (N+1)}
\nonumber\\ &&
+\,{64 \over 3 (N+2)^3}
-56\,{\frac {{ \zeta_2}}{{(N+1)}^{2}}}
-92\,{\frac {{ \zeta_2}}{(N+1)}}
-{\frac {32}{3}}\,{\frac {{ \zeta_2}}{(N+2)}}
+{\frac {104}{3}}\,{\frac {{ \zeta_2}}{(N-1)}}
\nonumber\\ &&
+{\frac {64}{3}}\,{ \frac {{ \beta^{(1)}} ( N+3 ) }{(N+2)}}
+{\frac {208}{3}}\,{\frac {{ \beta^{(1)}} ( N ) }{(N-1)}}
-16\,{\frac {{ \beta^{(2)}} ( N+2 ) }{(N+1)}}
-64\,{ \frac {{ A_3} ( N+1 ) }{(N+1)}}
\nonumber\\ &&
+64\,{\frac {{ \beta} ( N+2) { \zeta_2}}{(N+1)}}
+32\,{\frac {{ \zeta_3}}{(N+1)}}
-{\frac {128}{3}}\,{ \frac {{ S_3} ( N+1 ) }{(N+1)}}
-64\,{\frac {{ \beta^{(1)}} ( N+2) }{{(N+1)}^{2}}}
\nonumber\\ &&
-16\,{\frac {{ A_{18}} ( N+1 ) }{(N+1)}}
+{\frac { 256}{3}}\,{\frac {{ S_3} ( N ) }{N}}
+{\frac {20}{3}}\,{\frac {{ S_1^3} ( N+1 ) }{(N+1)}}
-{\frac {40}{3}}\,{\frac {{ S_1^3} ( N ) }{N} }
-32\,{\frac {{ \zeta_2}}{{(N-1)}^{2}}}
\nonumber\\ &&
+80\,{\frac {{ \zeta_3}}{(N-1)}}
-32\,{\frac { { S_3} ( N-1 ) }{(N-1)}}
-64\,{\frac {{ \beta^{(1)}} ( N ) }{{(N-1)}^{2}}}
-16\,{\frac {{ \beta^{(2)}} ( N ) }{(N-1)}}
-64\,{\frac {{ A_3} ( N-1 ) }{(N-1)}}
\nonumber\\ &&
+64\,{\frac {{ \beta} ( N ) { \zeta_2}}{(N-1)}}
+ \Bigg( {\frac {98}{3}}
+96\,{ \zeta_2}
-64\,{ \zeta_3}
-32\,{ \beta^{(2)}} ( N+1) 
+64\,{ \beta^{(1)}} ( N+1 ) 
\nonumber\\ &&
-128\,{ A_3} ( N ) 
+32\,{ A_{18}} ( N ) 
+128\,{ \beta} ( N+1 ) { \zeta_2} \Bigg) {1 \over N}
+ \Bigg( -{\frac {94}{3}}
+48\,{ \zeta_2} \Bigg) {1 \over N^2}
\nonumber\\ &&
+ \Bigg( \,{52 \over (N+1)^2}
+\,{86 \over (N+1)} \Bigg) { S_2} ( N+1 ) 
+ \Bigg(  \Bigg( -\,{68 \over (N+1)}
\nonumber\\ &&
-\,{24 \over (N+1)^2} \Bigg) { S_1} ( N+1 ) 
+ \Bigg( \,{80 \over N}
+\,{48 \over N^2} \Bigg) { S_1} ( N ) 
-16\,{\frac {{ S_1^2} ( N+1 ) }{(N+1)}}
+32\,{\frac {{ S_1^2} ( N ) }{N}}
\nonumber\\ &&
-16\,{\frac {{ S_2} ( N+1) }{(N+1)}}
+32\,{\frac {{ S_2} ( N ) }{N}}
+\,{28 \over (N+1)^2}
+32\,{ \frac {{ \zeta_2}}{(N+1)}}
-\,{52 \over (N+1)}
\nonumber\\ &&
+\,{24 \over (N+1)^3}
+{\frac {-64\,{ \zeta_2} +36}{N}}
-\,{64 \over N^2}
-\,{48 \over N^3} \Bigg) \log\left({Q^2 \over \mu^2}\right )
+ \Bigg( -\,{104 \over N^2}
\nonumber
\end{eqnarray}\begin{eqnarray}
%\nonumber\\ &&
&&
-\,{104 \over N} \Bigg) { S_2} ( N ) 
+ \Bigg( \,{20 \over (N+1)^2}
+\,{54 \over (N+1)}\Bigg) { S_1^2} ( N+1 ) 
+ \Bigg( -\,{40 \over N^2}
\nonumber\\ &&
-\,{72 \over N} \Bigg) { S_1^2} ( N ) 
+ \Bigg( 32\,{\frac {{ \zeta_2}}{(N-1)}}
+64\,{\frac {{ \beta^{(1)}} ( N ) }{(N-1)}}
-\,{16 \over (N-1)} \Bigg) { S_1} ( N-1 ) 
\nonumber\\ &&
+ \Bigg( \,{6 \over (N+1)}
-\,{4 \over (N+1)^2}
-16\,{\frac {{ S_1} ( N ) }{N}}
+8 \,{\frac {{ S_1} ( N+1 ) }{(N+1)}}
+\,{8 \over N^2} \Bigg)  \log^2\left({Q^2 \over \mu^2}\right )
\nonumber\\ &&
+ \Bigg( -56\,{\frac {{ S_2} ( N ) }{N}}
+{\frac {128\,{ \beta^{(1)}} ( N+ 1 ) -32\,{ \zeta_2}-88}{N}}
-\,{64 \over N^2}
-\,{16 \over N^3} \Bigg) { S_1} ( N) 
\nonumber\\ &&
+ \Bigg( 16\,{\frac {{ \zeta_2}}{(N+1)}}
+\,{52 \over (N+1)^2}
+\,{108 \over (N+1)}
+ 64\,{\frac {{ \beta^{(1)}} ( N+2 ) }{(N+1)}}
+28\,{\frac {{ S_2} ( N+1) }{(N+1)}}
\nonumber\\ &&
+\,{8 \over (N+1)^3} \Bigg) { S_1} ( N+1 ) 
\Bigg]
\end{eqnarray}
\end{appendix}

\newpage 
%%%%%%%%%%%%%%%%%%%%%%%%%%%%%%%%%%%%%%%%%%%%%%%%%%%%%%%%%%%%%%%%%%%%%%%%%% 
 
%%%%%%%%%%%%%%%%%%%%%%%%%%%%%%%%%%%%%%%%%%%%%%%%%%%%%%%%%%%%%%%%%%%%%%%%%%%%%%%%%%%%%%%%
\end{document}